\documentclass[aps,twocolumn,showpacs,preprintnumbers,nofootinbib,prd,superscriptaddress,groupedaddress,10pt]{revtex4-1}

\makeatletter
\def\l@subsubsection#1#2{}
\def\l@subsubsubsection#1#2{}
\makeatother

\usepackage{graphicx,amssymb,amsmath,amsthm,amsfonts,epsfig,epsf}
\usepackage[usenames]{color}
\usepackage{epstopdf}

\usepackage{aas_macros}
\usepackage{bm}
\usepackage{dcolumn}
\usepackage{latexsym}
\usepackage{rotating}
\usepackage{longtable}

\usepackage{enumerate}
\usepackage{tensor,multirow}
\usepackage{url}
\usepackage[linktocpage]{hyperref}

\def\nn{\nonumber}

\DeclareMathOperator{\sech}{sech}

\begin{document}

\title{An analytical template for gravitational-wave echoes: signal characterization and prospects of detection with current and future interferometers}
%%%%
\author{Adriano Testa}\email{adriano.testa@roma1.infn.it}
\affiliation{Dipartimento di Fisica, ``Sapienza'' Universit\`a di Roma \& Sezione INFN Roma1, Piazzale Aldo Moro 5, 00185, Roma, Italy}
\author{Paolo Pani}\email{paolo.pani@roma1.infn.it}
\affiliation{Dipartimento di Fisica, ``Sapienza'' Universit\`a di Roma \& Sezione INFN Roma1, Piazzale Aldo Moro 5, 00185, Roma, Italy}

\begin{abstract} 
Gravitational-wave echoes in the post-merger ringdown phase are under intense 
scrutiny as probes of near-horizon quantum structures and as signatures of 
exotic states of matter in ultracompact stars.
We present an analytical template that describes the ringdown and the echo 
signal for nonspinning objects in terms of two physical parameters: the 
reflectivity and the redshift at the surface of the object. We 
characterize the properties of the template and adopt it in a preliminary 
parameter estimation with current (aLIGO) and future (Cosmic Explorer, Einstein 
Telescope, LISA) gravitational-wave detectors. 
For fixed signal-to-noise ratio in the post-merger phase, the constraints on 
the model parameters depend only mildly on the details of the detector 
sensitivity 
curve, but depend strongly on the reflectivity. Our analysis suggests 
that it might be possible to detect or rule out Planckian corrections at the 
horizon scale for perfectly-reflecting ultracompact objects at $5\sigma$ 
confidence level with Advanced LIGO/Virgo. 
On the other hand, signal-to-noise ratios 
in the ringdown phase equal to $\approx 100$ (as achievable with future 
interferometers) might allow us to probe near-horizon quantum structures with 
reflectivity $\gtrsim30\%$ ($\gtrsim85\%$) at $2\sigma$ 
($3\sigma$) level. 
\end{abstract}

\maketitle

%%%%%%%%%%%%%%%%%%%%%%%%%%%%%%%%%%%%%%%%%%%%%%%%%%
\section{Introduction}
%%%%%%%%%%%%%%%%%%%%%%%%%%%%%%%%%%%%%%%%%%%%%%%%%%
Gravitational-wave~(GW) echoes in the post-merger GW signal from a binary
coalescence might be a generic feature of quantum corrections at the horizon 
scale~\cite{Cardoso:2016rao,Cardoso:2016oxy}, and might provide a smoking-gun 
signature of exotic compact objects~(ECOs) and of exotic states of matter in 
ultracompact stars~\cite{Ferrari:2000sr,Pani:2018flj} (for a review, 
see~\cite{Cardoso:2017cqb,Cardoso:2017njb}).
In the last two years, tentative evidence for echoes in the combined LIGO/Virgo 
binary black-hole~(BH) events have been 
reported~\cite{Abedi:2016hgu,Conklin:2017lwb} with controversial 
results~\cite{Ashton:2016xff,Abedi:2017isz,Westerweck:2017hus,Abedi:2018pst}. 
Recently, a tentative detection of echoes in the post-merger signal of 
neutron-star binary coalescence GW170817~\cite{TheLIGOScientific:2017qsa} has 
been claimed at $4.2\sigma$ confidence level~\cite{Abedi:2018npz}, but 
such a strong claim is yet to be confirmed/disproved by an independent 
analysis. The 
stochastic background produced by ``echoing remnants''~\cite{Du:2018cmp} and 
spinning ECOs~\cite{Barausse:2018vdb} has been also studied recently.

While model-independent~\cite{Abedi:2018npz} and burst~\cite{Tsang:2018uie} 
searches can be performed without knowing the details of the echo waveform, the 
possibility of extracting as much information as possible from post-merger 
events relies on one's ability to model the signal accurately. Furthermore, 
using an accurate template is crucial for model selection and to discriminate 
the origin of the echoes in case of a detection.
In the last year, there has been considerable progress in modeling the echo 
waveform~\cite{Nakano:2017fvh,Mark:2017dnq,Maselli:2017tfq,Bueno:2017hyj,
Wang:2018mlp,Correia:2018apm,Wang:2018gin}, but the proposed approaches are 
sub-optimal, because either they are based on analytical templates not 
necessarily 
related to the physical properties and parameters of the remnant, or they rely 
on numerical waveforms which are inconvenient for direct searches through 
matched filters.

In this work, we take the first step to overcome these limitations by building 
an \emph{analytical} template directly anchored to the physical properties of a 
given ECO model. 
As we shall discuss, the template captures the rich phenomenology of the GW 
echo signal, including amplitude and frequency modulation, which arise from the 
physical origin of the echoes, namely radiation that bounces back and forth 
between the object and the 
photon-sphere~\cite{Cardoso:2014sna}, slowly 
leaking to infinity through wave 
tunneling~\cite{Cardoso:2016rao,Cardoso:2016oxy,Cardoso:2017cqb,Cardoso:2017njb}
.

\begin{figure*}[th]
\centering
\includegraphics[width=0.475\textwidth]{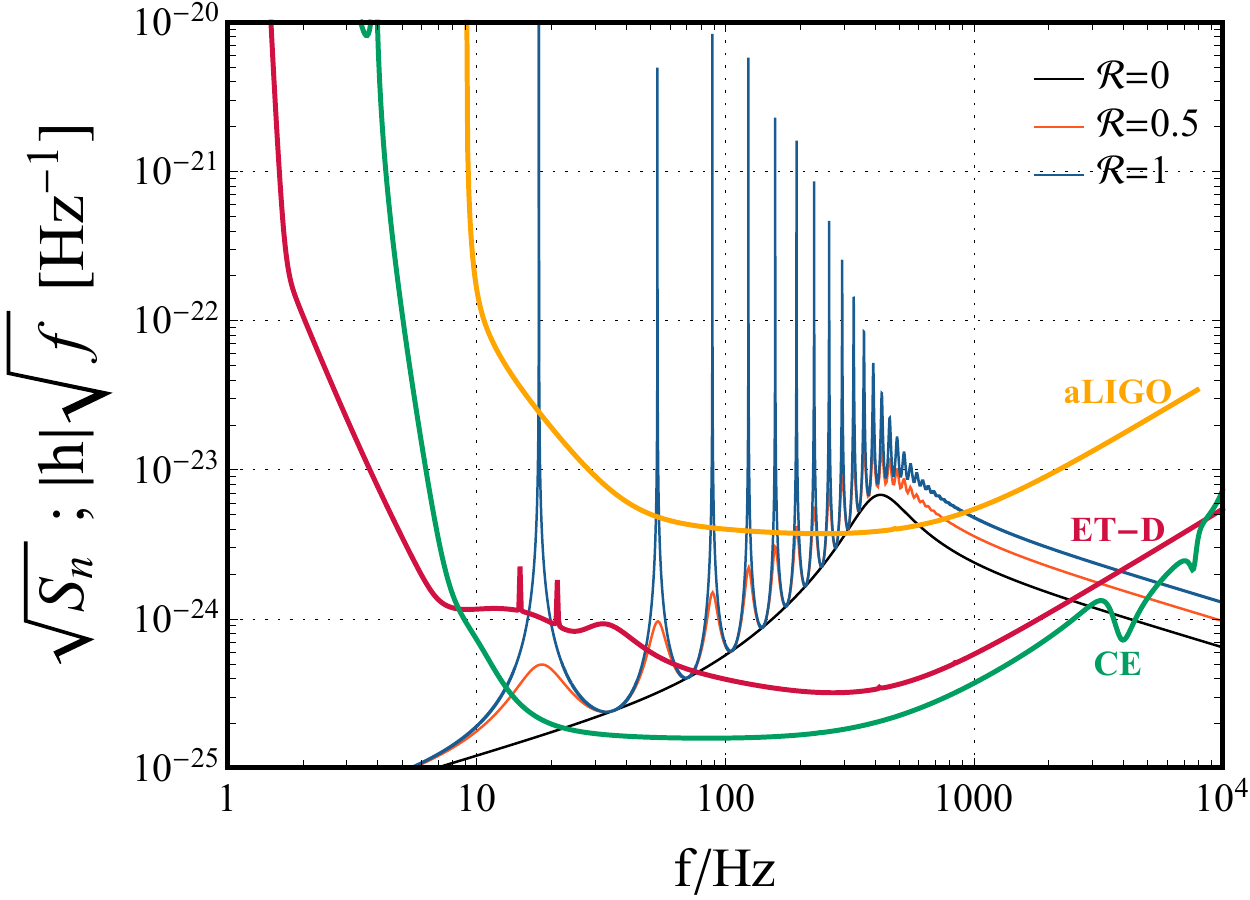}
\includegraphics[width=0.5\textwidth]{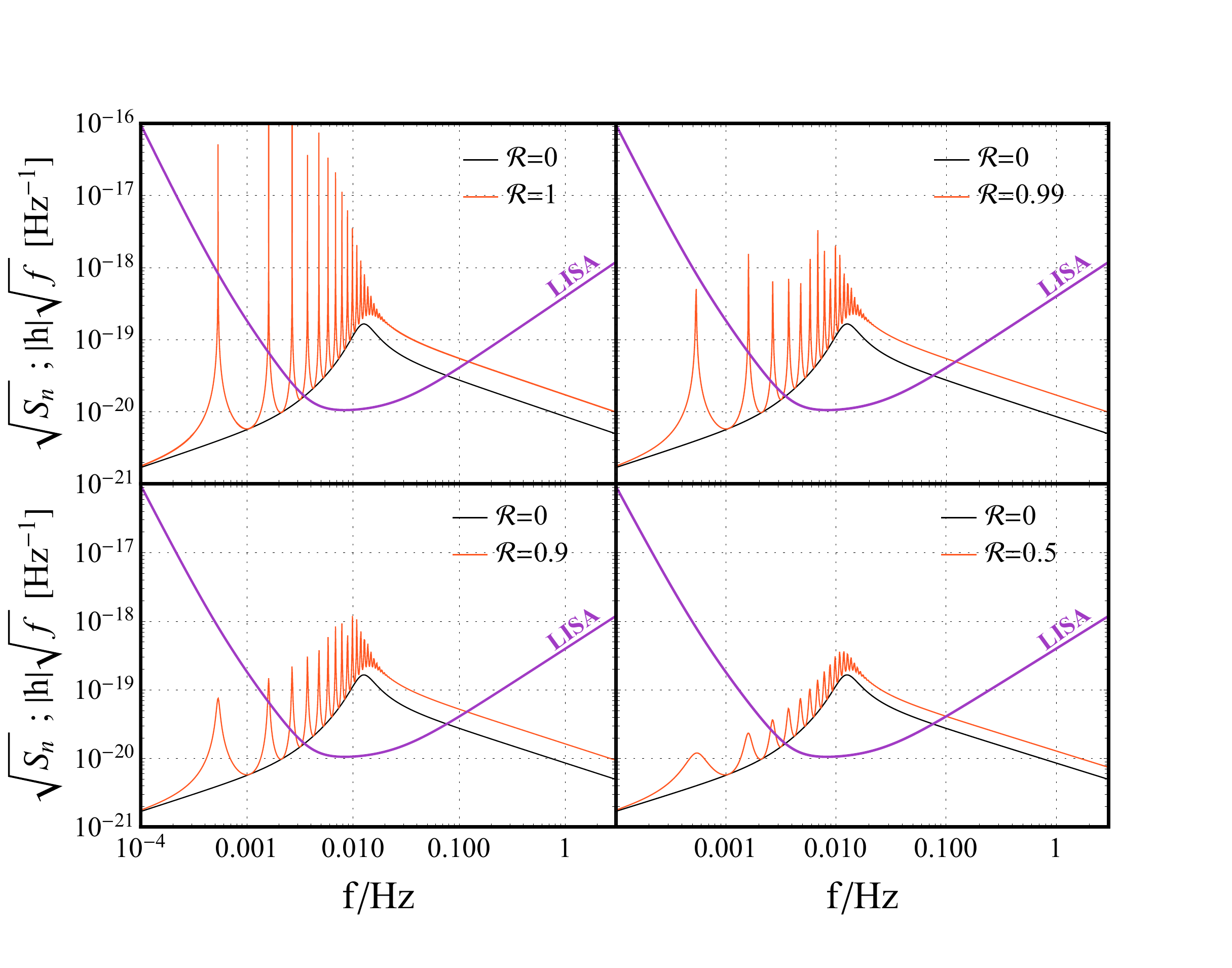}
\caption{Illustrative comparison between the ringdown$+$echo signal derived in 
this work and the 
power spectral densities of various interferometers as functions of the GW 
frequency $f$. Left panel: we 
considered an object with $M=30M_\odot$, at a distance of $D=400\,{\rm Mpc}$, 
with compactness parameter $d=100M$ (roughly corresponding to near-horizon 
quantum corrections, see below) and various values of the reflectivity 
coefficient ${\cal R}$ 
(${\cal R}=0$ corresponds to the pure BH ringdown template). The sensitivity 
curves 
refer to aLIGO with its anticipated design-sensitivity 
\texttt{ZERO\_DET\_high\_P} configuration~\cite{zerodet}, Cosmic~Explorer in 
the 
narrow band variant~\cite{Evans:2016mbw,Essick:2017wyl}, and Einstein Telescope 
in its \texttt{ET-D} configuration~\cite{Hild:2010id}. 
Right panels: the echo signal is compared to the recently proposed LISA's noise 
spectral density~\cite{LISA}. We considered an object with $M=10^{6}M_\odot$, 
at 
a distance of $D=100\,{\rm Gpc}$ (corresponding to cosmological 
redshift $\approx 10$), and with $d=100M$. Each small panel corresponds to a 
different 
value of ${\cal R}$. 
For simplicity in all panels we neglected corrections due to the geometry of 
the detector, sky averaging, and cosmological effects, and we assumed ${\cal 
A}\sim M/D$ for the amplitude of the ringdown signal. Details are given in the 
main text.} \label{fig:SNR}
\end{figure*}

As an illustration and anticipation of our results, in Fig.~\ref{fig:SNR} we 
compare the ringdown$+$echo signal derived below against the power spectral 
densities of current (aLIGO at design sensitivity~\cite{zerodet}) and future 
(Einstein~Telescope~\cite{Punturo:2010zz}, 
Cosmic~Explorer~\cite{PhysRevD.91.082001}, LISA~\cite{LISA}) 
GW interferometers. Details on the template are provided in Sec.~\ref{sec:Setup}.
A preliminary parameter 
estimation using current and future GW detectors is performed in 
Sec.~\ref{sec:bounds}.
We conclude in Sec.~\ref{sec:Discussion} with future prospects.
Throughout this work, we use $G=c=1$ units.

%%%%%%%%%%%%%%%%%%%%%%%%%%%%%%%%%%%%%%%%%%%%%%%%%%
\section{Setup}~\label{sec:Setup}
%%%%%%%%%%%%%%%%%%%%%%%%%%%%%%%%%%%%%%%%%%%%%%%%%%
As a first step, we focus on nonspinning models, the extension to spinning 
objects is underway.
Our approach is based on the analytical approximation of perturbations of the 
Schwarzschild geometry in terms of the P\"oschl-Teller 
potential~\cite{Poschl:1933zz,Ferrari:1984zz} and on the framework developed in 
Ref.~\cite{Mark:2017dnq}, in which the echo signal is written in terms of a 
transfer function that reprocesses the BH response at the horizon.
For the busy reader, our final template is provided in a ready-to-be-used form 
in Eq.~\eqref{FINALTEMPLATE} and in a supplemental {\scshape 
Mathematica}\textsuperscript{\textregistered} notebook~\cite{webpage}.

%%%%%%%%%%%%%%%%%%%%%%%%%
\subsection{An analytical template for GW echoes}
%%%%%%%%%%%%%%%%%%%%%%%%%

\begin{figure}[th]
\centering
\includegraphics[width=0.49\textwidth]{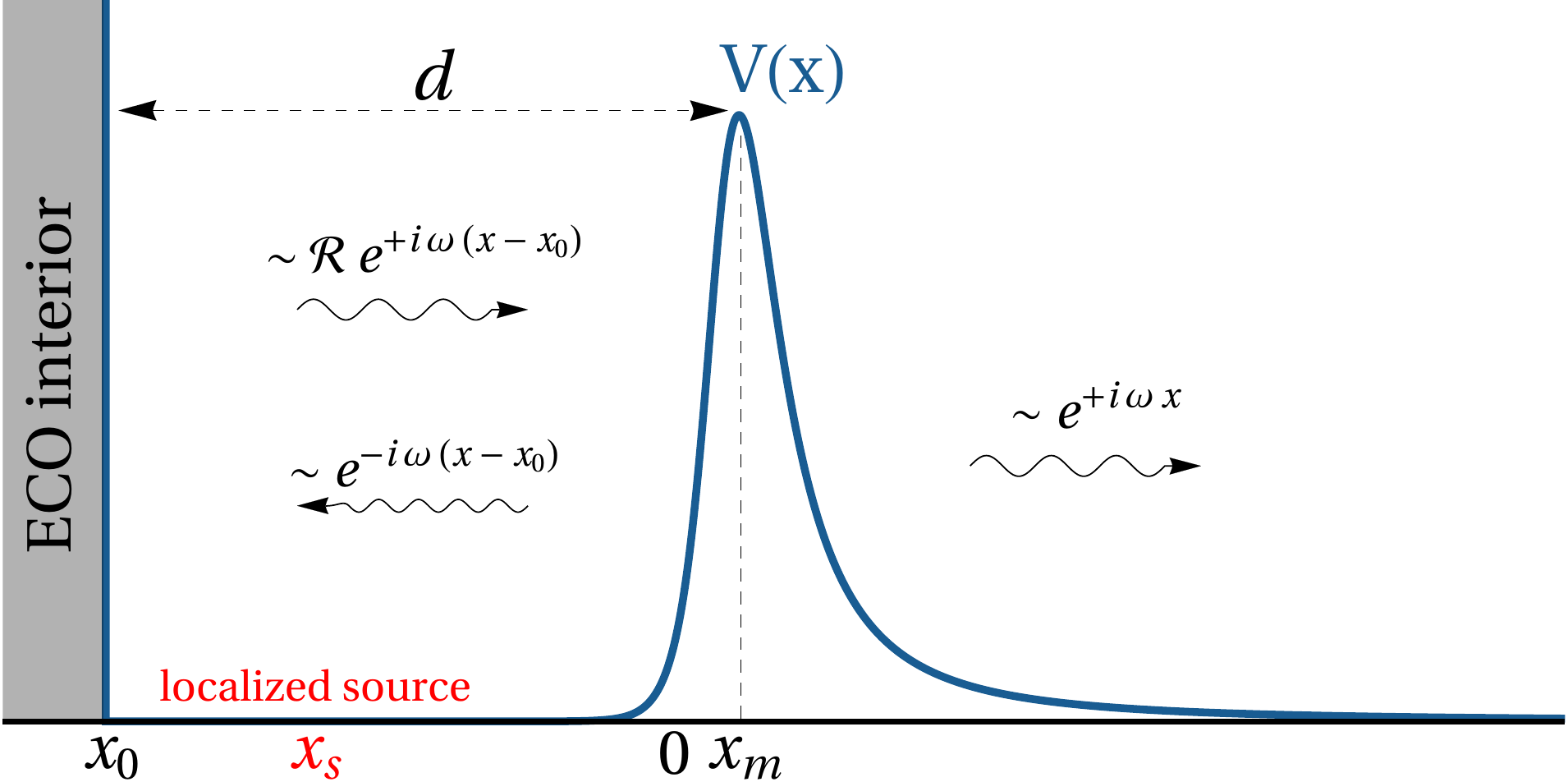}
\caption{Schematic description of our ECO model (adapted from 
Ref.~\cite{Mark:2017dnq}).}\label{fig:setting}
\end{figure}

We model the stationary ECO (see Fig.~\ref{fig:setting}) with a background 
geometry described, when $r>r_0$, 
by the Schwarzschild metric, 
\begin{equation}
 ds^2=-Adt^2+A^{-1}dr^2+r^2d\Omega^2\,,
\end{equation}
%%%
with $A(r)=1-2M/r$, $M$ and $r_0$ being the total mass and the radius of the 
object in Schwarzschild coordinates, respectively. At $r=r_0$, we assume the 
presence of a membrane, the properties of which are parametrized by a complex 
and (generically) frequency-dependent reflectivity 
coefficient ${\cal R}$~\cite{Mark:2017dnq,Maggio:2017ivp}.

%%%%%%%%%%%%%%%%%%%%%%%%%%%%%%%%
%%%%%%%%%%%%%%%%%%%%%%%%%%%%%%%%

This model is well suited to 
describe near-horizon quantum structures (which belong to the \emph{ClePhO} 
category introduced in Refs.~\cite{Cardoso:2017cqb,Cardoso:2017njb}).

\begin{figure}[th]
\centering
\includegraphics[width=0.49\textwidth]{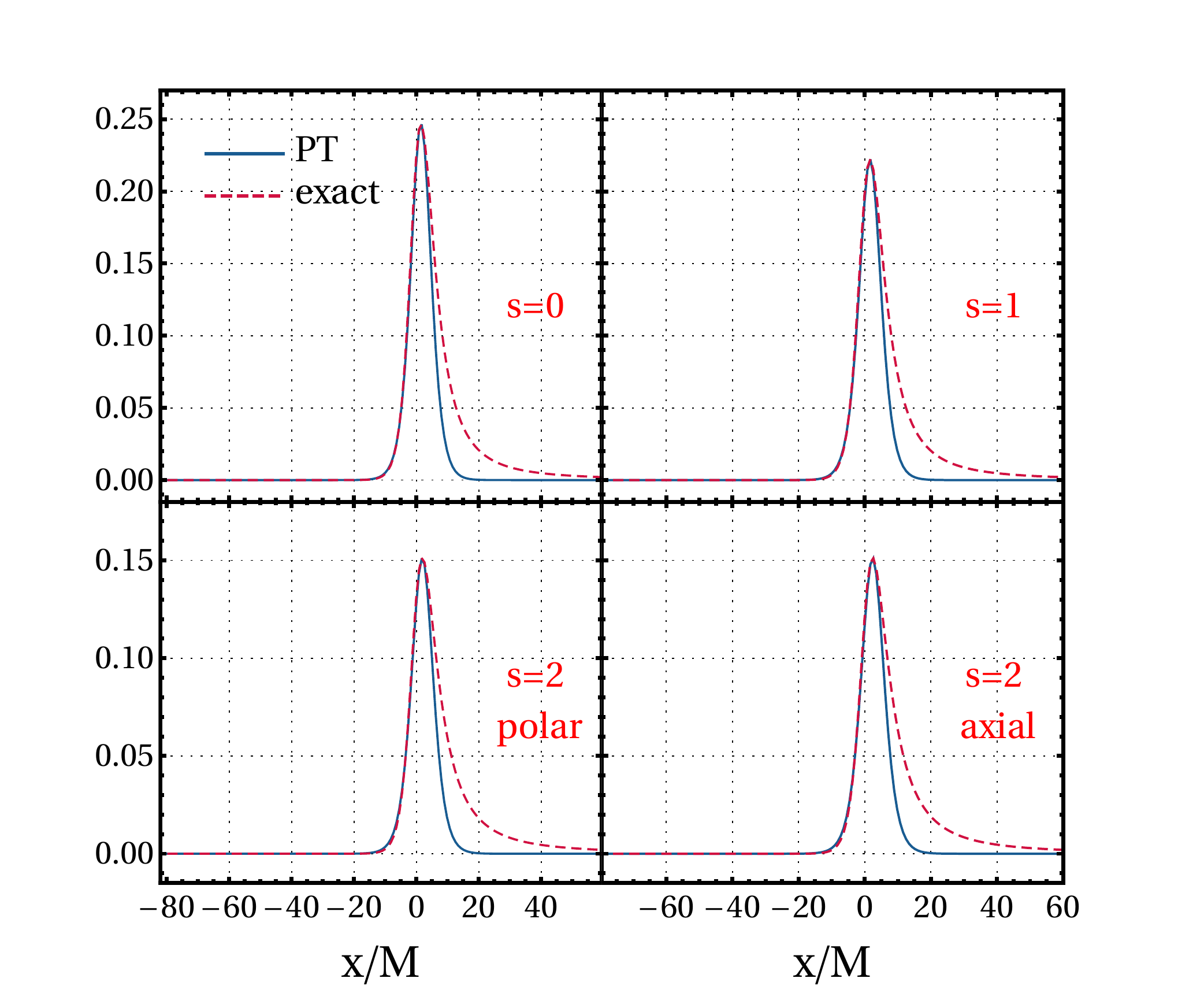}
\caption{Comparison between the exact potential governing perturbations in a 
Schwarzschild geometry (red dashed curve) and its approximation given by the 
P\"{o}schl-Teller potential (PT, blue continuous curve), as explained in the 
text.}\label{fig:PT}
\end{figure}

After carrying out a Fourier transform and a spherical-harmonics decomposition, 
various classes of perturbations of the background metric are described by a 
master equation
\begin{equation}
\left[\frac{\partial^2}{\partial 
x^2}+\omega^2-V_{sl}(r)\right]\tilde\Psi(\omega,x)=\tilde S(\omega,x) 
\,,\label{master}
\end{equation}
where
\begin{equation}\label{eq:tortoise}
x = r  + 2 M \log{ \left(\frac{r}{2 M} - 1 \right)}
\end{equation}
defines\footnote{We note that our definition of $x$ differs by a constant 
term $-2M\log2$ relative to the one adopted in Ref.~\cite{Mark:2017dnq}. 
\label{coordinate}} the tortoise coordinate $x$ of the Schwarzschild metric, 
$l$ 
is the multipolar index, $s$ identifies the type of the perturbation, and 
$\tilde S$ is a source term. 
We assume that the surface of the object in tortoise coordinates is located near 
the would-be horizon at $x_0=x(r_0)\ll - M$, as expected for near-horizon 
quantum 
corrections~\cite{Cardoso:2016rao,Cardoso:2016oxy,Cardoso:2017cqb,
Cardoso:2017njb}.
The potential reads
\begin{eqnarray}
V_{sl}(r)&=&A(r)\left(\frac{l(l+1)}{r^2}+\frac{1-s^2}{r}A'(r)\right)\,,
\label{potentialmaster}
\end{eqnarray}
where the prime denotes derivative with respect to the coordinate $r$. In the 
above potential, $l\geq s$ with $s=0,1$ for test 
Klein-Gordon and Maxwell fields, respectively, whereas $s=2$ for axial 
gravitational perturbations (see Fig.~\ref{fig:PT}). Also polar gravitational 
perturbations are described by Eq.~\eqref{master}, but in this case the 
potential reads
\begin{eqnarray}
V_{2l}^{\rm P}(r)&=&2A\left[\frac{9 M^3+9 M^2 r \Lambda +3 M r^2 \Lambda ^2+r^3 
\Lambda ^2
   (1+\Lambda )}{r^3 (3 M+r \Lambda )^2}\right]\,,\nn\\\label{potentialPolar}
\end{eqnarray}
with $\Lambda=(l-1)(l+2)/2$ and $l\geq2$.
While axial and polar perturbations of a Schwarzschild BH are 
isospectral~\cite{Chandra}, this property is generically broken for 
ECOs~\cite{Cardoso:2017cqb,Cardoso:2017njb}.

\subsubsection{Transfer function}

By using Green's function techniques, Mark et al.~\cite{Mark:2017dnq} showed 
that the solution of Eq.~\eqref{master} at infinity reads $\tilde 
\Psi(\omega,x\to\infty)\sim \tilde Z^{+}(\omega) e^{i\omega x}$, with
%%%%
\begin{equation}
 \tilde Z^{+}(\omega) = \tilde Z_{\rm BH}^{+}(\omega)+ {\cal K}(\omega) \tilde 
Z_{\rm BH}^{-}(\omega)\,. \label{signalomega}
\end{equation}
%%%%
In the above equation, $\tilde Z_{\rm BH}^{\pm}$ are the responses of a 
Schwarzschild BH (at infinity and near the horizon, for the plus and minus 
signs, respectively) to the source $\tilde S$,
%%%
\begin{equation}
 \tilde Z_{\rm BH}^{\pm}(\omega) = \int_{-\infty}^{+\infty}dx \frac{\tilde S 
\tilde \Psi_{\mp} }{W_{\rm BH}}\,, \label{ZBH}
\end{equation}
%%%
where $W_{\rm BH}=\frac{d\tilde \Psi_+}{dx}\tilde \Psi_- -\tilde 
\Psi_+ \frac{d\tilde \Psi_-}{dx}$ is the Wronskian, and $\tilde \Psi_\pm$ are 
the solutions of the homogeneous equation associated 
to Eq.~\eqref{master} such that
\begin{equation}\label{Psip}
\tilde \Psi_+(\omega, x) \sim \begin{cases}
 \displaystyle 
e^{+i \omega x} & \text{ as } x \to + \infty\\ 
 \displaystyle  
 B_{\rm out}(\omega)e^{+i \omega x}  +  B_{\rm in}(\omega) e^{- i \omega x} & \text{ as } x \to - \infty
\end{cases} \,,\\
\end{equation}
\begin{equation}
\tilde \Psi_-(\omega, x) \sim \begin{cases}
 \displaystyle 
 A_{\rm out}(\omega)e^{+i \omega x}  +  A_{\rm in}(\omega) e^{-i \omega x} & \text{ as } x \to + \infty \\ 
 \displaystyle  
 e^{-i \omega x} & \text{ as } x \to - \infty \\
\end{cases} \, .
\end{equation}
The details of the ECO model are all contained in the transfer function
%%%%
\begin{equation}
{\cal K}(\omega)=\frac{{\cal T}_{\rm BH} {\cal R} e^{-2 i  \omega x_0}}{1-{\cal 
R}_{\rm BH} {\cal R} e^{-2 i  \omega x_0 }}\,, \label{transfer}
\end{equation}
%%%%
where ${\cal T}_{\rm BH} = 1/B_{\rm out}$ and ${\cal R}_{\rm BH} = B_{\rm 
in}/B_{\rm out}$ are the transmission and reflection coefficients for waves 
coming from the \emph{left} of 
the BH potential barrier~\cite{Mark:2017dnq,Chandra,NovikovFrolov}, whereas 
${\cal R}(\omega)$ is the reflection coefficient at the surface of the object, 
defined so that
%%%
\begin{equation}\label{eq:boundary}
 \tilde \Psi\to e^{-i\omega (x- x_0)} + {\cal R}(\omega) e^{i\omega (x- x_0)}\,,
\end{equation}
%%%
near the surface at $x\sim x_0$, with $|x_0|\gg M$. In 
Appendix~\ref{app:optics}, we provide a heuristic derivation of 
Eq.~\eqref{transfer} in terms of a geometrical optics analogy.
%%%
The above equations are subject to the constraint that the time domain 
waveforms 
are real, which implies
%%%
\begin{equation}\label{eq:kreal}
{\cal K}(\omega) = {\cal K}^*(-\omega^*)\,
\end{equation}
%%%
and analogous relations for the other quantities.

We notice that the signal in the frequency domain is written as a sum of the BH 
response plus an extra piece proportional to $\cal R$. The poles of this latter 
piece are the quasinormal modes (QNMs) of the ECO which, in general, differ 
from those of the BH. Since the BH QNMs scale with $1/M$ while the ECO QNMs 
scale with $1/d$, when $d\gg M$ the signal at small times is 
dominated by the poles of the BH (which are no longer QNMs since they do not 
satisfy the right boundary conditions, see discussion below).

\begin{table*}[th]
 \begin{tabular}{c|ccc|cccc}
 \hline
 \hline
 Potential				&$s$	&$\omega_RM$&$\omega_IM$& 
$\alpha M$	& $V_0 M^2$		& $x_m/M$    & $ 
{\Delta\omega_I}/{\omega_I}$\\
 \hline
 scalar 				&$0$	&0.4836& -0.09676& $0.2298$	
& $0.2471$	& $1.466$ 	& $0.1876$ \\
 electromagnetic 		&$1$	&0.4576& -0.09500& $0.2265$	& 
$0.2222$	& $1.614$		& $0.1921$ \\
 axial gravitational 		&$2$	&0.3737& -0.08896& $0.2159$	& 
$0.1513$	& $2.389$ 	& $0.2136$\\
 polar gravitational 		&$2$	&0.3737& -0.08896& $0.2161$	& 
$0.1513$	& $1.901$ 	& $0.2148$\\
 \hline
 \hline
 \end{tabular}	
 \caption{Numerical values of the fitting parameters of the P\"{o}schl-Teller 
potential~\eqref{PT} used in this work to approximate the exact potential [see 
Fig.~\ref{fig:PT}]. Scalar, electromagnetic and axial gravitational 
perturbations are described by the potential~\eqref{potentialmaster}, whereas 
polar gravitational perturbations are described by the 
potential~\eqref{potentialPolar}. We restrict to $l=2$. As an indicator of the 
quality of the analytical approximation, in the last column we show the 
relative 
difference $\frac{\Delta\omega_I}{\omega_I}$ between the exact imaginary part 
of 
the frequency (shown in the fourth column) and that given by the 
P\"{o}schl-Teller potential. The parameter 
$x_m$ is expressed in terms of the tortoise coordinate defined in 
Eq.~\eqref{eq:tortoise}.
}\label{tab:PT}
\end{table*}

%%%%%
\subsubsection{P\"{o}schl-Teller potential}
%%%%
The potential~\eqref{potentialmaster} [and~\eqref{potentialPolar}] can be 
approximated by the P\"{o}schl-Teller 
potential~\cite{Poschl:1933zz,Ferrari:1984zz,Bueno:2017hyj}
%%%
\begin{equation}
V_{{\rm PT}}(x) = \frac{V_0}{\cosh^2 [\alpha(x - x_m)]}\,, \label{PT}
\end{equation}
%%%%
where $\alpha$, $V_0$ and $x_m$ are free parameters. We chose $V_0$ and $x_m$ 
such that the position of the maximum and its value coincide with those of the 
corresponding $V_{sl}$. The remaining parameter $\alpha$ was found by 
imposing that the real part of the fundamental QNM of the 
P\"{o}schl-Teller potential,
%%%%
\begin{equation}
\omega_R = \sqrt{V_0 - \alpha^2/4}\,,
\end{equation}
%%%
coincides with the exact one, as found numerically~\cite{Berti:2009kk}.

The values of $\alpha$, $V_0$ and $x_m$ obtained through this procedure for 
different classes of potentials are given in Table~\ref{tab:PT}. The 
P\"{o}schl-Teller potential provides an excellent approximation on the left of 
the potential barrier and near the maximum (see Fig.~\ref{fig:PT}), which are 
crucially the most relevant regions for our model. On the right of the potential 
barrier the behavior is different, since the P\"{o}schl-Teller potential decays 
exponentially as $x\gg M$, whereas the exact potential decays as $\sim 1/x^2$. 
We have checked that this different behavior would affect only the reprocessing 
of very low frequency signals, but it is negligible for the first echoes, which 
are characterized by the reprocessing of the dominant QNMs with $\omega_R M\sim 
1$.

%%%%

Using the approximate potential, the homogeneous equation corresponding to 
Eq.~\eqref{master} can be solved analytically. The general solution of the 
homogeneous problem, $\tilde\Psi_0$, can be expressed in terms of associated 
Legendre functions as
%%%
\begin{equation}
\tilde\Psi_0= c_1 P_{\frac{i \omega_R }{\alpha } - \frac{1}{2}}^{\frac{i \omega 
}{\alpha }}(2 \xi-1)+c_2 Q_{\frac{i \omega_R }{\alpha } - \frac{1}{2}}^{\frac{i
   \omega }{\alpha }}(2 \xi-1)\,, \label{tildePsi0}
\end{equation}
where $c_1$ and $c_2$ are integration constants and $\xi = [{1 + e^{-2 
\alpha(x - x_m)}}]^{-1}$ is a new variable.
%%%%

Taking $c_1 = e^{i \omega x_m} \Gamma(1 - \frac{i \omega}{\alpha})$ and $c_2 = 
0$, $\tilde\Psi_0$ reduces to $\tilde \Psi_+$
and we obtain
\begin{eqnarray}
{\cal T}_{\rm BH} &=& -\frac{i}{\pi} \sinh   \left(\frac{\pi  \omega }{\alpha 
}\right)\Upsilon\,,  \label{TBH}\\
{\cal R}_{\rm BH} &=&-\frac{1}{\pi }\cosh\left(\frac{\pi 
\omega_R}{\alpha}\right) \Upsilon e^{2 i \omega x_m} \,, \label{RBH}
\end{eqnarray}
where we defined
%%%
\begin{equation}
 \Upsilon=\Gamma\left(\frac{1}{2}  - 
i\frac{\omega+\omega_R}{\alpha}\right)\Gamma\left(\frac{1}{2}  - 
i\frac{\omega-\omega_R}{\alpha}\right)\frac{\Gamma\left(1+\frac{i\omega}{\alpha}
\right)}{\Gamma\left(1-\frac{i\omega}{\alpha}\right)}\,.\label{gamma}
\end{equation}
By replacing the above expressions in Eq.~\eqref{transfer}, we finally obtain
%%%
\begin{equation}
 {\cal K}(\omega)= -i\frac{e^{2i\omega d}{\cal 
R}(\omega)\Upsilon\sinh{\left(\frac{\pi  \omega}{\alpha 
}\right)}}{\pi+e^{2i\omega d}{\cal R}(\omega)\cosh{(\frac{\pi 
\omega_R}{\alpha})}\Upsilon}e^{-2i\omega x_m}\,, \label{Kappa}
\end{equation}
%%%
where we defined the width of the cavity\footnote{For ECOs with near-horizon 
quantum structures, one expects $d\sim nM|\log(l_P/M)|$, where $l_P$ is the 
Planck length and $n\sim {\cal O}(1)$ depends on the 
model~\cite{Cardoso:2016oxy,Cardoso:2017cqb,Cardoso:2017njb}. This gives 
$d/M\sim 100n$ roughly for both stellar-mass and supermassive objects.
In this 
case, the redshift at the surface roughly reads $z\sim M/l_P$.}
$d=x_m-x_0>0$ (recall that $x_0<0$ and $x_m>0$), which is also related to the 
redshift at the surface, $z\sim e^{d/(4M)}$ when $d\gg M$. 
As we shall show later, the final signal depends only on the physical quantity 
$d$; the dependence on $x_m$ in Eq.~\eqref{Kappa} will disappear from the final 
result\footnote{This must be the case and, in fact, it is possible to add a 
phase term in the definitions of the transfer function and of the BH response 
at 
the horizon so that $x_m$ never appears in the equations. We chose not to do 
so in order to follow the notation of Ref.~\cite{Mark:2017dnq} more closely. 
Equivalently, in a 
coordinate system such that the maximum of the potential sits at the origin, 
all results would depend only on the physical quantity $d$.}.

In summary, for a given choice of ${\cal R}(\omega)$, the above relations yield 
an \emph{analytical} approximation to the transfer function ${\cal K}$. In 
Appendix~\ref{app:comparison}, we compare some results for the 
approximate analytical expressions of ${\cal T}_{\rm BH}$, ${\cal R}_{\rm BH}$ 
and ${\cal K}$, with their exact numerical counterparts as computed in 
Ref.~\cite{Mark:2017dnq}.

%%%%%%%%%%%%%%%%%%%%%%%%%%%%%%%%%%
\subsubsection{Modeling the BH response}
%%%%%%%%%%%%%%%%%%%%%%%%%%%%%%%%%%
The inverse Fourier transform of the BH response $\tilde Z_{\rm 
BH}^\pm(\omega)$ 
[see Eq.~\eqref{ZBH}] can be deformed in the complex frequency plane, yielding 
three contributions~\cite{Leaver:1986gd,Berti:2009kk}: (i) the high-frequency 
arcs that govern the prompt response; (ii) a sum-over-residues at the poles of 
the complex frequency plane (defined by $W_{\rm BH}=0$), which correspond to 
the 
QNMs and that dominate the signal at intermediate times; (iii) a branch cut on 
the negative half of the imaginary axis, giving rise to late-time tails due to 
backscattering 
off the background curvature.
The post-merger ringdown signal is very well approximated by the second 
contribution only, so that for most astrophysical applications the BH response 
at infinity can be written as a superposition of QNMs~\cite{Berti:2009kk}. 
Considering for simplicity only the dominant mode, one gets
%%%
\begin{equation}
 Z_{\rm BH}^{+} (t)\sim \mathcal{A}\, \theta(t - t_0)  \cos(\omega_R t+\phi)  
e^{-t/\tau}\,, \label{ZBHplus}
\end{equation}
%%%
where the complex QNM frequency reads $\omega_R+i\omega_I$, $\tau=-1/\omega_I$, 
${\cal A}$ and $\phi$ are the amplitude and the phase, respectively, and $t_0$ 
parametrizes the starting time of the ringdown. In the above expression we have 
defined ${\cal A}= M {\cal A}_{lmn} S_{lmn}/D$, where $M$ is the mass of the 
object, $D$ is the distance of the source, $A_{lmn}$ is the amplitude of the 
BH QNM with quantum numbers $(l,m,n)=(2,2,0)$, and $S_{lmn}$ are the 
corresponding
spin-weighted spheroidal harmonics~\cite{Berti:2005ys}. 
%%%
Given the BH response in the time domain, the frequency-domain waveform is
obtained through the Fourier transform,
\begin{equation}
\tilde Z_{\rm BH}^{\pm}(\omega) = \int_{- \infty}^{+ \infty} \frac{dt}{\sqrt{2 
\pi}} Z_{\rm BH}^{\pm}(t) e^{i \omega t}.
\end{equation}
%%%
For the BH response at infinity, the Fourier transform yields
%%%
%%%
\begin{equation}\label{eq:bhtemplateINF}
 \tilde Z_{\rm BH}^{+}(\omega) \sim \frac{{\cal A}}{2\sqrt{2\pi}}  \left( 
\frac{\alpha_1}{\omega - \Omega_+} +  \frac{\alpha_2}{\omega - \Omega_-} 
\right)e^{i \omega t_0},
\end{equation}
%%% 
where $\Omega_\pm = \pm \omega_R + i \omega_I$, $\alpha_1=-ie^{-i(\phi+t_0 
\Omega_+)}$ and $\alpha_2=-{\alpha_1}^*$.

It is worth noting that the complex poles are the same for $Z_{\rm BH}^+$ and 
for $Z_{\rm BH}^-$, since those are defined by $W_{\rm BH}=0$. This suggests 
that the BH response near the horizon might also be described by a 
superposition 
of QNMs\footnote{Note that, since the P\"oschl-Teller potential is symmetric 
under reflections around its maximum, within our approximation the BH response 
near the horizon would be equivalent to the response at infinity provided the 
source term in Eq.~\eqref{ZBH} has the same symmetry of the potential. However, 
this is generally not the case.\label{notasimmetria}}.

%%%%%
\begin{table}[th]
 \begin{tabular}{ll}
 \hline
 \hline
  $M$			& total mass of the object \\
  ${\cal A}$		& amplitude of the BH ringdown  \\
  $\phi$		& phase of the BH ringdown  \\
  $t_0$			& starting time of the BH ringdown \\ 
  \hline
  $d$  			& width of the cavity ($z\sim e^{d/(4M)}$) \\
  ${\cal R}(\omega)$	& reflection coefficient at the surface \\ 
  \hline
  \hline
 \end{tabular}
 \caption{Parameters of the ringdown$+$echo template presented in this work. The 
first four parameters characterize the ordinary BH ringdown. The parameter $z$ 
is the gravitational redshift at the ECO surface.} \label{tab:template}
\end{table}
%%%
%%%%
More in general, we are interested in perturbations produced by sources 
localized near the object, as expected for merger remnants. If we assume $\tilde 
S(\omega,x)=C(\omega) \delta(x-x_s)$,
for any $x_s$ well inside the cavity (where $V(x_s)\approx0$), from 
Eq.~\eqref{ZBH} we can derive a relation between $\tilde 
Z_{\rm BH}^{+}$ and $\tilde Z_{\rm BH}^{-}$, namely:
%%%
\begin{eqnarray}
 \tilde Z_{\rm BH}^- = e^{2i\omega x_s}\left(1+{\cal R}_{\rm BH}e^{-2i\omega 
x_s} \right)\frac{\tilde  Z_{\rm BH}^+}{{\cal T}_{\rm BH}}\,. 
\label{eq:bhtemplateHOR}\\ \nn 
\end{eqnarray}
%%%
Remarkably, the above relation is independent of the function $C(\omega)$ 
characterizing the source.
Thus, $\tilde Z_{\rm BH}^-$ can be computed analytically using the expressions 
for ${\cal R}_{\rm BH}$, ${\cal T}_{\rm BH}$ and $\tilde  Z_{\rm BH}^+$ derived 
above. As 
expected, the quantity ${\cal K} \tilde Z_{\rm BH}^-$ [appearing in 
Eq.~\eqref{signalomega}] is independent of $x_m$ and depends only on the 
physical width of the cavity $d$.

An expression similar to Eq.~\eqref{eq:bhtemplateHOR} can be obtained for a 
source localized\footnote{Another particular case is 
when the 
source is 
localized well outside the 
light ring ($x_s\gg M$). In this case we obtain
\begin{equation}\label{greenout}
 \tilde Z_{\rm BH}^+ = e^{-2i\omega x_s}\left(1+\hat{\cal R}_{\rm 
BH}e^{2i\omega x_s} \right)\frac{\tilde  Z_{\rm BH}^-}{\hat{\cal T}_{\rm BH}}\,,
\end{equation}
where $\hat{\cal R}_{\rm BH}$ and $\hat{\cal T}_{\rm BH}$ are respectively the 
reflection and transmission coefficients for the scattering of left-moving
waves 
\emph{from} infinity. Note that, since the P\"oschl-Teller 
potential is symmetric around its maximum, within our framework these 
coefficients coincide with ${\cal R}_{\rm BH}$ and ${\cal T}_{\rm BH}$, 
respectively, modulo a phase difference.} at any point $x_s$ by 
using the explicit form of 
Eq.~\eqref{tildePsi0}~\cite{thesis}. The generic expression is given in 
Appendix~\ref{app:genericsource}.
In what follows we will consider a source localized near the surface, $x_s 
\approx x_0$, which should provide a model for post-merger excitations.
%

%%%%%%%%%%%%%%%%%
\subsubsection{Ringdown$+$echo template}
%%%%%%%%%%%%%%%%%
Putting together all the ingredients previously derived, the final expression 
of 
the full ECO response reads
%%%
% \begin{widetext}
 \begin{equation}
\tilde Z^+(\omega) = Z_{\rm BH}^+(\omega) \left[1+{\cal R}\frac{\pi-e^{2 i 
\omega d}\Upsilon\cosh\left(\frac{\pi \omega_R}{\alpha}\right)}{\pi+e^{2 i 
\omega d}{\cal R}\Upsilon\cosh\left(\frac{\pi \omega_R}{\alpha}\right)} \right] 
\,, 
\label{FINALTEMPLATEGEN}
 \end{equation}
% \end{widetext}
%%%
where $\Upsilon$ is defined in Eq.~\eqref{gamma}.
In this form, the signal is written as the reprocessing of a generic BH response 
$Z_{\rm BH}^+(\omega)$. Note that the BH response $Z_{\rm BH}^+(\omega)$ might 
not necessarily be restricted to a ringdown signal in the frequency domain. In 
general, if the remnant is an ECO, one might expect that the post-merger phase 
can be obtained through the reprocessing of the ringdown part \emph{and} of the 
late-merger phase, since already during the formation of the final ECO radiation 
might be trapped by an effective photon-sphere.

If instead we model the BH response $Z_{\rm BH}^+(\omega)$ through a single 
ringdown [Eq.~\eqref{eq:bhtemplateINF}], the explicit final form of the 
ringdown$+$echo 
signal reads
%%%
\begin{widetext}
 \begin{equation}
  \tilde Z^+(\omega) = \sqrt{\frac{\pi}{2}}{\cal A} \frac{  
e^{i(\omega-\omega_I)t_0} (1+{\cal R})  \Gamma\left(1-\frac{i \omega }{\alpha
   }\right) (\omega_R\sin(\omega_R t_0+\phi)+i (\omega +i \omega_I) 
\cos(\omega_R t_0+\phi)}{\left[(\omega +i \omega_I)^2-\omega_R^2\right]
   \left[\pi  \Gamma\left(1-\frac{i \omega }{\alpha }\right)+e^{2 i d \omega } 
{\cal R}
   \cosh\left(\frac{\pi \omega_R}{\alpha}\right)\Gamma\left(\frac{1}{2}  - 
i\frac{\omega+\omega_R}{\alpha}\right) \Gamma\left(\frac{1}{2}  - 
i\frac{\omega-\omega_R}{\alpha}\right)
   \Gamma\left(1+\frac{i \omega }{\alpha }\right)\right]} \,. 
\label{FINALTEMPLATE}
 \end{equation}
\end{widetext}
%%%
The above result is valid for any type of perturbation, provided the parameters 
$\alpha$ and $V_0$ (or, equivalently, $\alpha$ and $\omega_R$) are chosen 
appropriately (see Table~\ref{tab:PT}). In the next section we will use 
Eq.~\eqref{FINALTEMPLATE} for polar gravitational perturbations.

For a linearly polarized wave, our template is defined by $4$ real parameters 
(${\cal A}$, $\phi$, $t_0$ and $d$) plus the mass (which sets the scale for the 
other dimensionful quantities) and the complex function ${\cal R}(\omega)$ that 
is model dependent (see Table~\ref{tab:template}). Clearly, for ${\cal R}=0$ 
one 
recovers a single-mode BH ringdown template in the frequency domain. The echo 
contribution is fully determined only in terms of $d$ and ${\cal R}(\omega)$ 
once the parameters of the ordinary ringdown are known.  If two polarizations 
are included, the number of real parameters increases to $7$ (${\cal A}_+$, 
${\cal A}_\times$, $\phi_+$, $\phi_\times$, $t_0$, $M$, $d$), plus the function
${\cal 
R}(\omega)$.

The templates~\eqref{FINALTEMPLATEGEN} and~\eqref{FINALTEMPLATE} are also 
publicly available in a ready-to-be-used supplemental {\scshape 
Mathematica}\textsuperscript{\textregistered} notebook~\cite{webpage}.

%%%%%%%%%%%%%%%%%%%%%%%%%
\subsection{Properties of the template}
%%%%%%%%%%%%%%%%%%%%%%%%%
\subsubsection{Comparison with the numerical result}

\begin{figure}[th]
\centering
\includegraphics[width=0.47\textwidth]{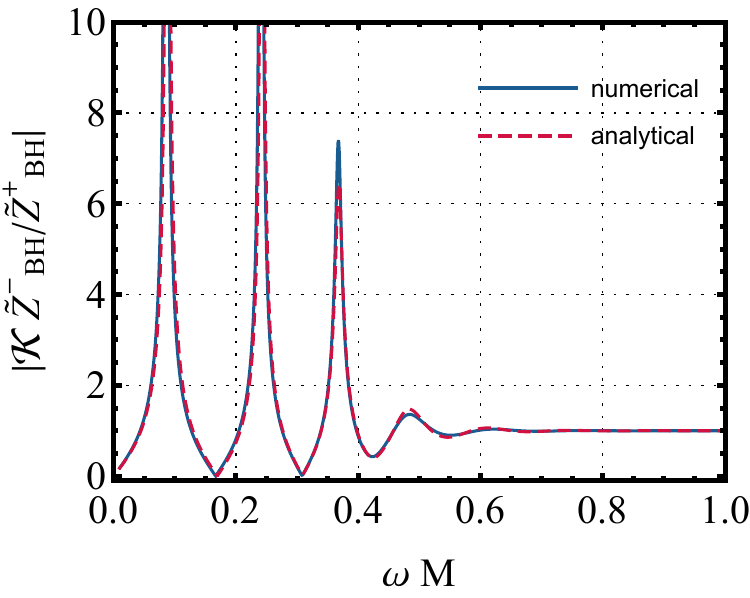}
\caption{Comparison between our analytical template and the result of a 
numerical integration of the Regge-Wheeler equation. We show the 
(absolute value of the) second term of the right hand side of Eq.~\eqref{signalomega}, normalized by 
the standard BH response, $\tilde Z_{\rm BH}^+$, for $d\approx 
20M$, $l=2$ axial perturbations, and a source localized near the ECO 
surface.} \label{fig:comparison_signal_FD}
\end{figure}

Our final analytical template agrees remarkably well with the exact numerical 
results. A representative example is shown in 
Fig.~\ref{fig:comparison_signal_FD}. Here we compare the template 
against the result of a 
numerical integration of the Regge-Wheeler equation for a source localized 
near the ECO surface. We show the second term of the right hand side of Eq.~\eqref{signalomega}, normalized by the standard BH response $Z_{\rm 
BH}^+$ (since $Z_{\rm 
BH}^-$ is proportional to $Z_{\rm BH}^+$, the final result is independent of the 
specific BH response). The agreement is overall very good and also the 
resonances are properly reproduced.

\subsubsection{Time-domain echo signal}
The time-domain signal can be simply computed through an inverse Fourier 
transform,
%%%
\begin{equation}
h(t) = \frac{1}{\sqrt{2 \pi}}\int_{- \infty}^{+ \infty} d\omega 
\tilde{Z}^{+}(\omega) e^{-i \omega t}.
\end{equation}
%%%%

\begin{figure*}[th]
\centering
\includegraphics[width=0.95\textwidth]{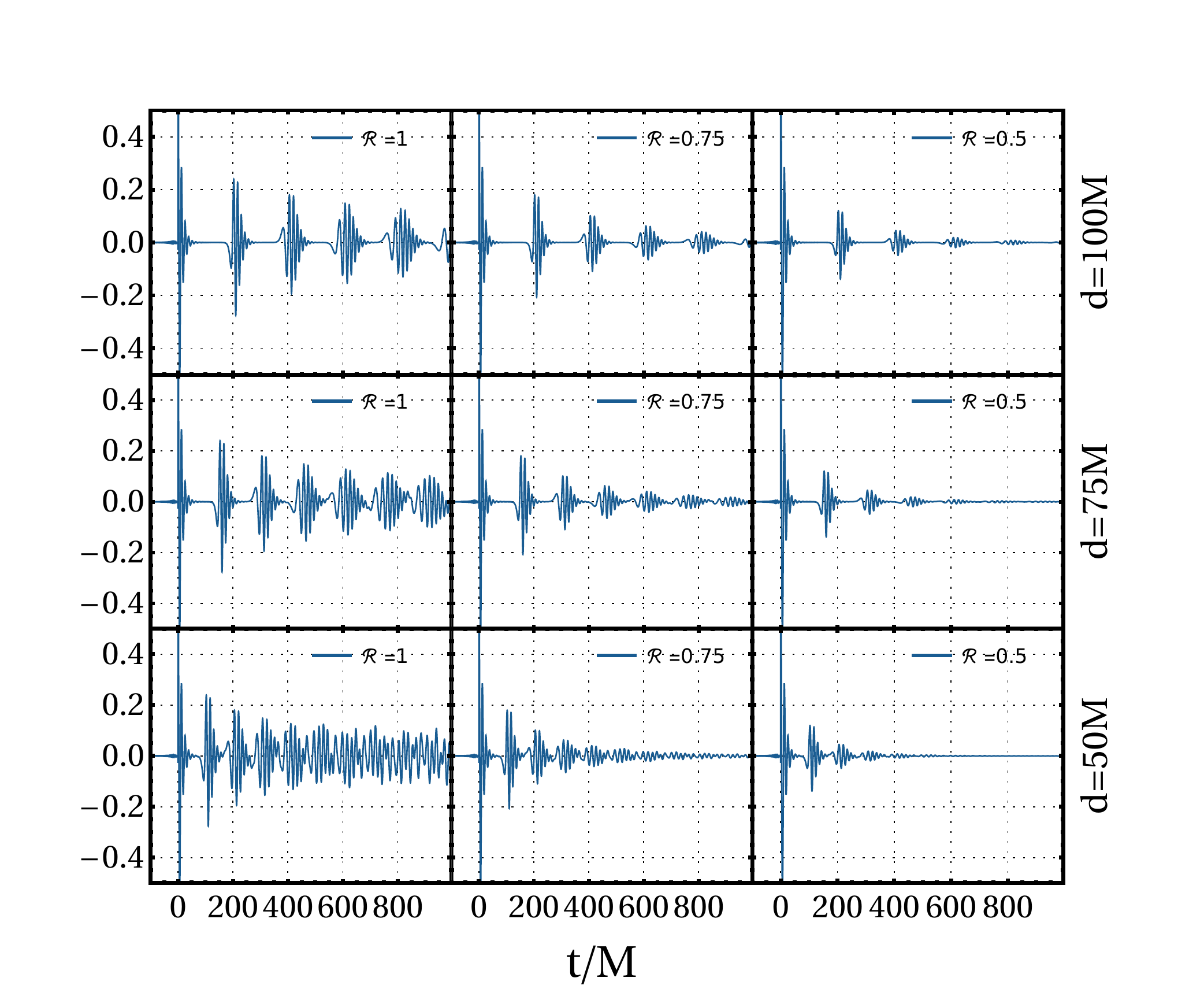}
\caption{Examples of the ringdown$+$echo template in the time domain for 
different values of $d$ and ${\cal R}(\omega)={\rm const}$ and for scalar 
perturbations. From top to bottom: $d=100M$, $d=75 M$, $d= 50 M$; from left to 
right: ${\cal R}=1$, ${\cal R}=0.75$, ${\cal R}=0.5$. The waveform is 
normalized 
to its peak value during the ringdown (not shown in the range of the $y$ axis 
to better visualize the subsequent echoes).} \label{fig:template0}
\end{figure*}

\begin{figure*}[th]
\centering
\includegraphics[width=0.95\textwidth]{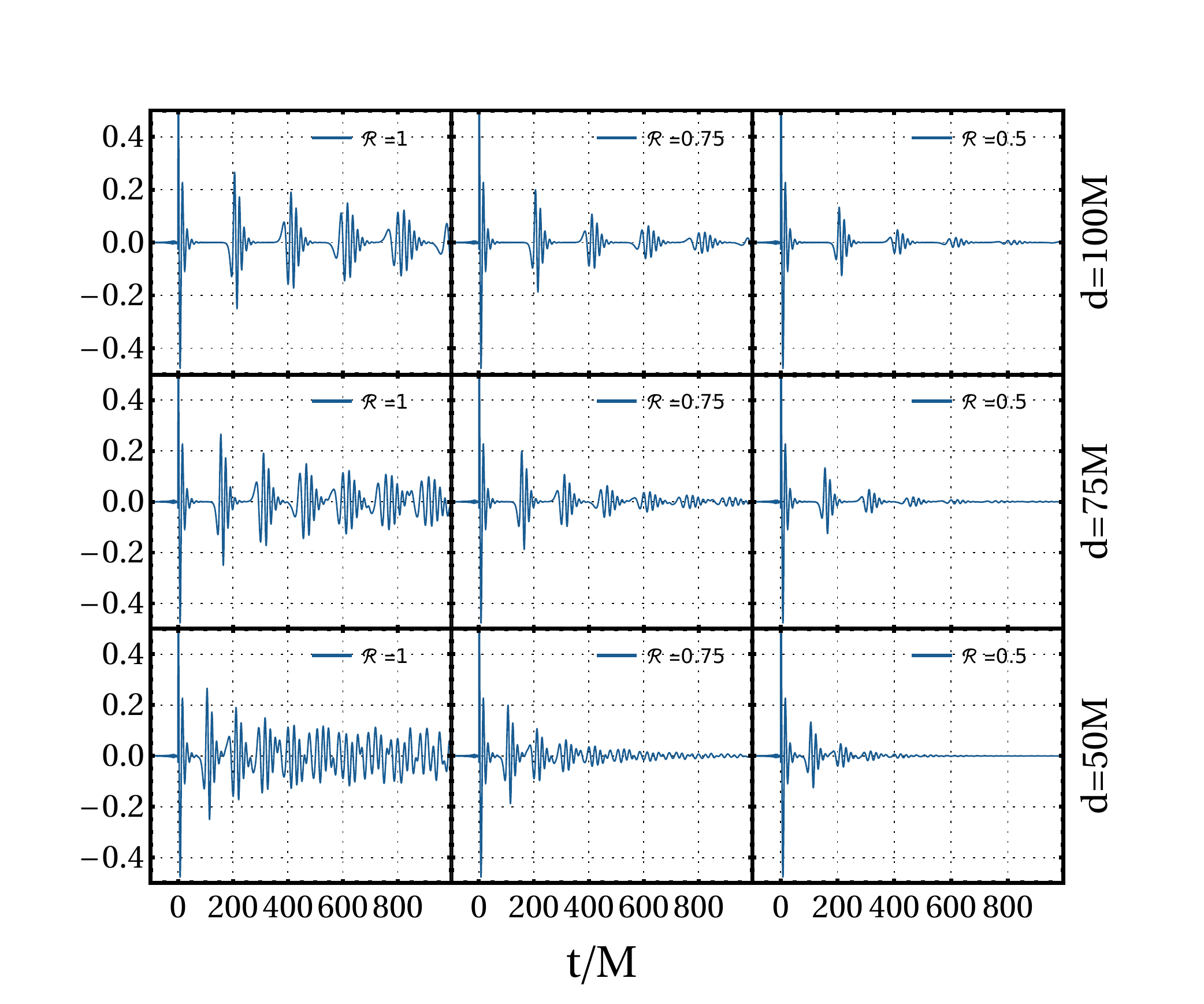}
\caption{Same as in Fig.~\ref{fig:template0}, but for polar gravitational 
perturbations.}
\label{fig:template2}
\end{figure*}

In Figs.~\ref{fig:template0} and~\ref{fig:template2}, we present a 
representative slideshow of our template for different values of $d$, ${\cal 
R}(\omega)$, and for scalar and polar gravitational perturbations, 
respectively. 
For simplicity, we consider ${\cal R}(\omega)={\rm const}$.
The time-domain waveform contains all the features previously reported 
for the echo signal, in particular amplitude and frequency 
modulation
\cite{Cardoso:2016rao,Cardoso:2016oxy,Cardoso:2017cqb,Cardoso:2017njb}, and 
phase inversion~\cite{Abedi:2016hgu,Wang:2018gin} of each echo relative to the 
previous one due to the reflective boundary conditions.
%%%%

%%%%%%%
%%%%%

\subsubsection{Decay of the echo amplitude in time}

\begin{figure}[th]
\centering
\includegraphics[width=0.52\textwidth]{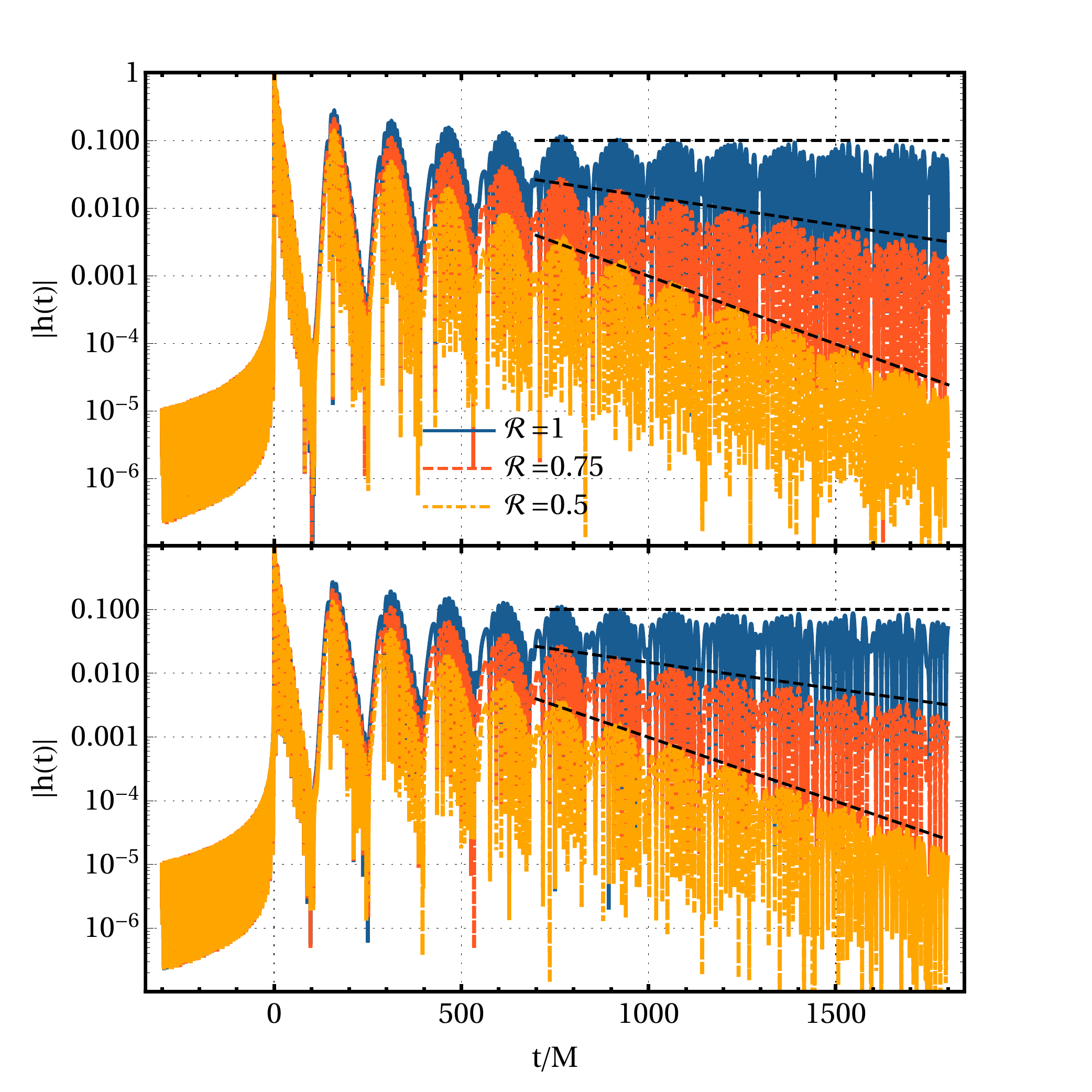}
\caption{Normalized absolute value of 
$h(t)$ for scalar (top panel) and polar gravitational (bottom panel) echo 
template. Continuous black lines show the slope $|h(t)|\sim 0.1|{\cal 
R}|^{t/(2d)}$, see Eq.~\eqref{slope}. We set $d=75M$ and considered 
${\cal R}=(1,0.75,0.5)$; different choices of the parameters give similar 
results.}
\label{fig:slope}
\end{figure}

Several qualitative features of the waveforms can be understood 
with a simple geometrical-optics toy model presented in 
Appendix~\ref{app:optics}. From this model, we expect the complex amplitude of 
each echo (in the frequency domain) relative to the previous one to be 
suppressed by a frequency-dependent factor ${\cal R}{\cal R}_{\rm BH}$, where 
we dropped 
the 
phase term $e^{- 2 i \omega d}$ that accounts for the time delay between the 
two. 
The first echo has already a factor ${\cal R}_{\rm BH}(\omega)$, which is 
essentially a low-pass filter. As shown in Fig.~\ref{fig:TBK}, ${\cal R}_{\rm 
BH}(\omega) \approx 0$ for $\omega \gtrsim \omega_R$, whereas $|{\cal R}_{\rm 
BH}(\omega)| \approx 1$ for $\omega \lesssim \omega_R$. Thus, for frequencies 
$\omega<\omega_R$, we expect that the amplitude of the echoes in 
the time domain should decrease as
%%%
\begin{equation}
 A_{\rm echo}(t)\propto  |{\cal R}{\cal R}_{\rm BH}|^\frac{t}{2d} \approx |{\cal 
R}|^\frac{t}{2d} \,.\label{slope}
\end{equation}
%%%
As shown in Fig.~\ref{fig:slope}, 
Eq.~\eqref{slope} agrees almost perfectly with our numerical results in the time
domain (we expect the small departure of the line from the data to be a 
consequence of the fact that ${\cal R}_{\rm BH}$ and ${\cal T}_{\rm BH}$ 
are not exactly constant).

%%%%%%%%%%%%%%%%%%%%%%%%
%%%%%%%%%%%%%%%%%%%%%%%%

\subsubsection{Phase inversion of subsequent echoes}

The phase inversion between subsequent echoes shown in 
Figs.~\ref{fig:template0} and \ref{fig:template2} can be understood by 
considering the extra 
factor ${\cal R}{\cal R}_{\rm BH}$ each echo has with respect to the previous 
one. When ${\cal R}$ is real and positive, the phase is set by ${\cal R}_{\rm 
BH}(\omega)$. Since ${\cal R}_{\rm BH}(\omega \sim 0) \approx -1$ for 
low-frequency signals (which are the only ones that survive the first filtering 
by ${\cal R}_{\rm BH}$), the $n$-th echo has a phase factor $\left[{\cal 
R}_{\rm 
BH}(\omega \sim 0)\right]^n \approx (-1)^n$. If the ECO is a wormhole, there is 
no phase 
inversion because in this case ${\cal R}={\cal R}_{\rm BH} 
e^{-2i\omega 
x_0}$~\cite{Mark:2017dnq}, so that ${\cal R}{\cal R}_{\rm BH} = {\cal R}_{\rm 
BH}^2$, where again we dropped the time-shifting phase. 
Likewise, if ${\cal R}$ is real and negative, the $n$-th echo has a phase 
factor $\left[{\cal R}{\cal R}_{\rm 
BH}\right]^n \approx 1$ for any $n$, so also in this case there is no phase 
inversion. Although 
not shown in Figs.~\ref{fig:template0} and \ref{fig:template2}, we have checked 
that all these properties are reproduced by the time-domain templates. 
Additional waveforms are provided online~\cite{webpage}.

\subsubsection{Dependence on the location of the source}

From the geometrical optics analogy of Appendix~\ref{app:optics}, we expect 
that, for a source located inside the cavity at some $x_s = x_0 + {\ell}$, 
the 
effect of the surface will appear only after a (coordinate-time) delay of $2 
{\ell}$ with respect to the RD, because of the extra time it takes for the 
left-going perturbation to 
reach the surface and come back.
Since the latter has a relative amplitude ${\cal R}$, the amplitude of the 
prompt signal is 
$\approx{\cal A}(1+{\cal R} e^{2i\omega {\ell}})$.
If $x_s \approx x_0$ (i.e., $\ell\approx0$), there is no delay between the 
proper ringdown signal and the first reflection. This is consistent with the 
behavior of the signal shown in Fig.~\ref{fig:SNR}: the full response at high 
frequencies (i.e., those which are not reflected by the potential but only by 
the surface) differs from the BH ringdown by a relative factor $1+{\cal R}$. 
Note that, if $x_s=x_0$, in the Dirichlet case (${\cal R}=-1$) the prompt 
signal and the reflected one interfere with opposite phase and the signal 
vanishes, as clear from Eq.~\eqref{FINALTEMPLATE}.

When the source is located outside the light ring 
[and consequently one needs to use Eq.~\eqref{greenout}], the frequency 
content of the ECO response at infinity changes significantly. Indeed, in 
this 
case the low-frequency content of an incident packet would not be able to probe 
the cavity and would be reflected regardless of the nature of the object and of 
the boundary conditions at $x_0$. On the other hand, the very high frequency 
component should pass through the light ring barrier unmodified and be 
reflected only by the ECO surface.
The frequency-domain signal for a source localized at a generic location $x_s$ 
is given in Appendix~\ref{app:genericsource}.

\subsubsection{Energy of echo signal}

\begin{figure}[th]
\centering
\includegraphics[width=0.49\textwidth]{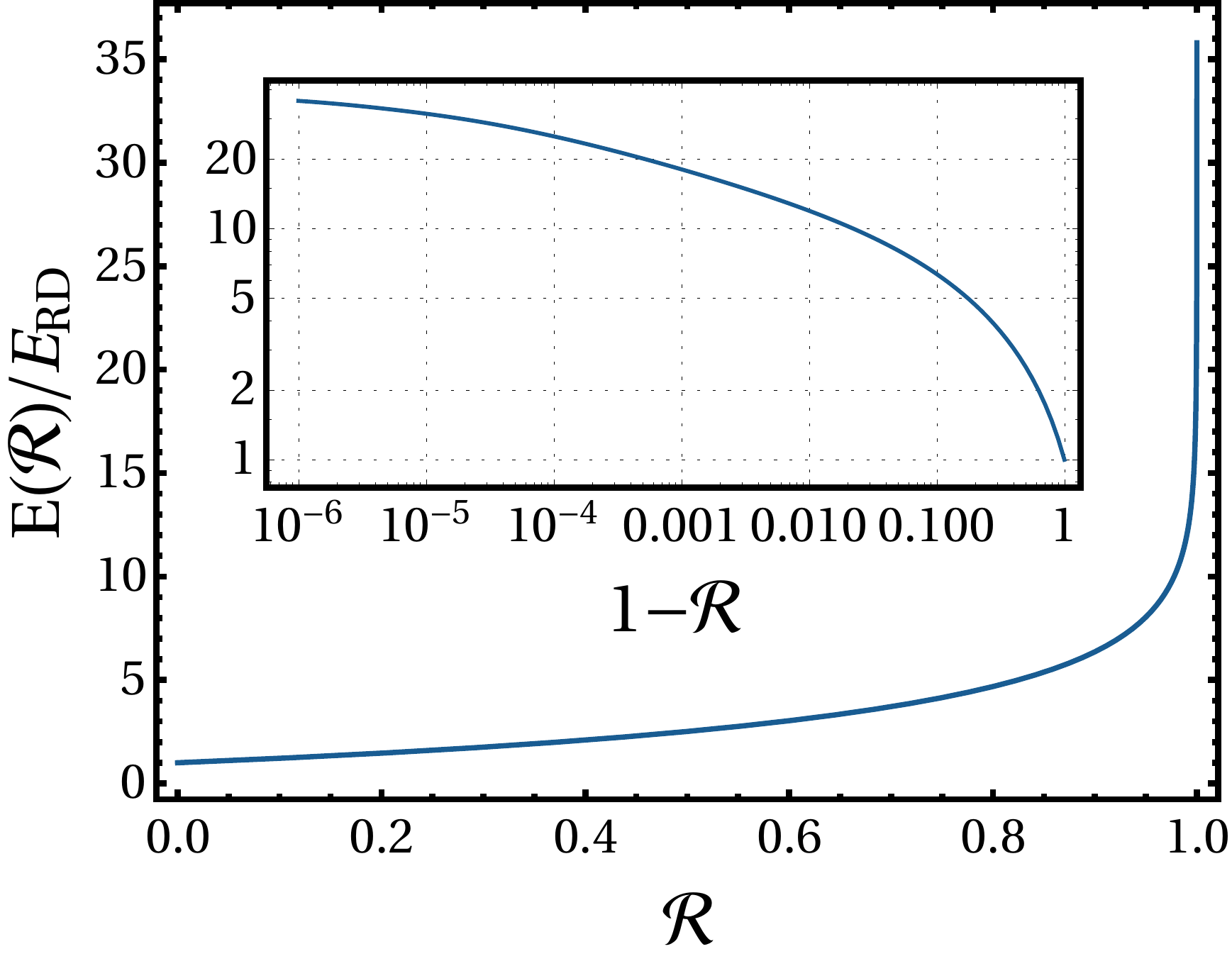}
\caption{Total energy contained in the ringdown$+$echo signal normalized by 
that 
of the ringdown alone as a function of $\cal {R}$.
The inset shows the same quantity as a function of $1-{\cal R}$ in a 
logarithmic 
scale.
Note that the energy is much larger than the ringdown energy when ${\cal 
R}\to1$. We set $t_0=0=\phi$ and $d=100M$; the result is independent of $d$ in 
the large-$d$ limit.} \label{fig:EvsR}
\end{figure}

Finally, let us discuss the energy contained in the ringdown$+$echo signal.
Because of partial reflection, the energy contained in the full signal is 
always 
larger 
than that of the ringdown itself (for ${\cal R} >0$). This is shown in 
Fig.~\ref{fig:EvsR}, where 
we 
plot 
the energy 
%%%
\begin{equation}
 E\propto \int_0^\infty d\omega \omega^2 |\hat Z^+|^2\,, \label{energy}
\end{equation}
%%%
normalized by the one 
corresponding to the ringdown alone, $E_{\rm RD}=E({\cal R}=0)$, as a function  
of the reflectivity ${\cal R}$. In the above equation, $\hat Z^+$ is the 
frequency-domain full response obtained by using the Fourier transform of 
%%%
\begin{equation}
 Z_{\rm BH}^{+} (t)\sim \mathcal{A}\, \cos(\omega_R t+\phi)  e^{-|t|/\tau}\,, 
\label{ZBHplusB}
\end{equation}
%%%
rather than using Eq.~\eqref{ZBHplus}. This is the prescription used in 
Ref.~\cite{Flanagan:1997sx} to compute the ringdown energy, and it circumvents 
the fact that the Heaviside function in Eq.~\eqref{ZBHplus} produces a spurious 
high-frequency behaviour of the energy flux, leading to infinite energy in the 
ringdown signal. With the above prescription, the energy defined in 
Eq.~\eqref{energy} is finite and reduces to the result of 
Ref.~\cite{Flanagan:1997sx} for the BH ringdown 
when ${\cal R}=0$.

As shown in Fig.~\ref{fig:EvsR}, the energy contained in the ringdown$+$echo 
signal depends strongly on ${\cal R}$ and can be $\approx 38$ times 
larger than that of the ringdown 
alone when ${\cal R}\to 1$ (the exact number depends also on $t_0$ and $\phi$).
This is due to the resonances\footnote{As shown in Fig.~\ref{fig:SNR}, when 
${\cal R}=1$ the resonances are very narrow and high. However, the most 
narrow resonances contribute little to the total energy. For example, in the 
extreme case ${\cal R}=1$ and $d=100M$, the integral~\eqref{energy} converges 
numerically when using fixed frequency steps $\Delta f\approx0.001/M$ or 
smaller. This suggests that not resolving all resonances in the signal might 
not 
lead to a significant loss in the SNR. We are grateful to Jing Ren, whose 
comment prompted this analysis.} corresponding to the low-frequency QNMs of the 
ECO, that can be excited with large amplitude [see Fig.~\ref{fig:SNR}]. At high 
frequency there are no resonances, but the energy flux $dE/d\omega$ is larger 
than the ringdown energy flux by a factor $(1+{\cal R})^2$, in agreement with 
our previous discussion.

%%%%%%%%%%%%%%%%%%%%%%%%%%%%%%%%%%%%%%%%%%
\subsubsection{Connection with the QNMs}
%%%%%%%%%%%%%%%%%%%%%%%%%%%%%%%%%%%%%%%%%%
The final expression~\eqref{FINALTEMPLATEGEN} also helps clarifying why the 
``prompt'' ringdown signal of an ECO is identical to that of a BH even if the 
BH QNMs are not part of the QNM spectrum of an 
ECO~\cite{Cardoso:2016rao,Barausse:2014tra}. While this fact is easy to 
understand in the time domain due to causality (in terms of time needed 
for the perturbation to 
probe the boundaries~\cite{Cardoso:2016rao}), it is less obvious in the 
frequency domain. 

The crucial point is that the BH QNMs are still poles of the ECO full 
response, $\tilde Z^+(\omega)$, in the complex frequency plane, even if they 
are not part of the ECO QNM spectrum. Indeed, Eq.~\eqref{FINALTEMPLATEGEN} 
contains two types of complex poles: those associated with $\tilde Z_{\rm 
BH}^+(\omega)$ which are the standard BH QNMs (but do not appear in the ECO QNM 
spectrum), and those associated with the poles of the transfer function ${\cal 
K}$, which correspond to the ECO QNMs. 
%%%
The late-time signal is dominanted by the second type of poles, since the 
latter have lower frequencies than the BH QNMs. On the other hand, the prompt 
ringdown is dominanted by the first type of poles, i.e. by the dominant QNMs of 
the corresponding BH spacetime.

%%%%%%%%%%%%%%%%%%%%%%%%%%%%%%%%%%%%%%%%%%%%%%%%%%
\section{Projected constraints on ECOs} \label{sec:bounds}
%%%%%%%%%%%%%%%%%%%%%%%%%%%%%%%%%%%%%%%%%%%%%%%%%%

In this section we use the template previously derived 
[Eq.~\eqref{FINALTEMPLATE}] for a preliminary parameter estimation of the ECO 
properties using current and future GW detectors. 
We shall focus on polar gravitational perturbations with $l=2$, which are 
typically the dominant ones.

As already illustrated in Fig.~\ref{fig:SNR}, the ringdown$+$echo signal 
displays sharp peaks which originate from the resonances of the transfer 
function ${\cal K}$ [see Fig.~\ref{fig:TBK}] and 
correspond to the long-lived QNMs of the ECO. Indeed, they are very well 
described by the harmonics of normal modes in a cavity of width $d$ 
[see Eq.~\eqref{modes}] (with a 
finite resonance width given by the small imaginary part of the 
mode~\cite{Cardoso:2014sna,Brito:2015oca}) and their frequency separation is 
$\Delta \omega\propto 1/d$. The 
relative amplitude of each resonance in the signal depends on the source, and 
the dominant modes are not necessarily the fundamental 
harmonics~\cite{Mark:2017dnq,Bueno:2017hyj}.

Although only indicative, Fig.~\ref{fig:SNR} already shows two important 
features. First, the amplitude of the echo signal in the frequency domain can 
be 
\emph{larger} than that of the ringdown itself. This explains the 
aforementioned 
energy content of the echo signal with respect to the ordinary ringdown 
(Fig.~\ref{fig:EvsR}) and suggests that GW echoes 
might be detectable even when the ringdown is not (we note that this feature is 
in agreement with some claims of 
Refs.~\cite{Abedi:2016hgu,Conklin:2017lwb,Abedi:2018npz}). 
Second, as shown in the right panels of Fig.~\ref{fig:SNR}, the amplitude of 
the echo signal depends strongly on the value of ${\cal R}$ and changes by 
several orders of magnitude between ${\cal R}=0.5$ and ${\cal R}=1$ (and 
even more for smaller values of ${\cal R}$). This suggests that the 
detectability of (or the constraints on) echoes will strongly depend on ${\cal 
R}$ and would be much more feasible when ${\cal R}\approx 1$. Below we shall 
quantify this expectation.

%%%%%%%%%%%%%
\subsection{Fisher analysis}
%%%%%%%%%%%%%

For simplicity, we employ a Fisher analysis, which is accurate at large 
signal-to-noise ratios (SNRs) (see, e.g., Ref.~\cite{Vallisneri:2007ev}). 
We shall assume that the signal 
is linearly polarized; including two polarizations is a 
straightforward extension.

Furthermore, since our ringdown$+$echo template was built for nonspinning 
ECOs, in 
principle we should also neglect the spin of the final object. However, since 
the statistical errors obtained from the Fisher matrix depend on the number of 
parameters, it is more realistic to assume that the template depends on the 
spin 
(through $\omega_R$ and $\omega_I$, taken to be those of a Kerr BH rather than 
of 
a Schwarzschild BH) and to perform the Fisher analysis by injecting a 
vanishing value of the spin. This procedure introduces some systematics, since 
we are ignoring the remaining spin dependence of the echo template. 
Nonetheless, 
it should provide a more reliable order-of-magnitude estimate of the 
statistical errors on the ECO parameters in the spinning case.

The Fisher matrix $\Gamma$ of a template $\tilde h(f)$ for a detector with 
noise 
spectral density $S_n(f)$ is defined as
%%%
\begin{equation}\label{fisher}
\Gamma_{i j} = (\partial_i \tilde h, \partial_j \tilde h)= 4 \, \Re 
\int_{0}^{\infty} \frac{\partial_i \tilde h^*(f) \partial_j \tilde 
h(f)}{S_n(f)} 
df\,,
\end{equation}
%%%
where $i,j=1,...,N$, with $N$ being the number of parameters in the template, 
and $f=\omega/(2\pi)$ is the GW frequency.
The SNR $\rho$ is defined as
\begin{equation}
\rho^2 = 4 \int_{0}^{\infty} \frac{\tilde h^*(f) \tilde h(f)}{S_n(f)} df\,. 
\label{SNR}
\end{equation}
The inverse of the Fisher matrix, $\Sigma_{ab}$, is the covariance matrix of 
the 
errors on the template's parameters: $\sigma_{i}=\sqrt{\Sigma_{ii}}$ gives the 
statistical error associated with the measurement of $i$-th parameter.

We note that, to the leading order in the large SNR limit, the statistical 
errors estimated through the Fisher matrix are independent of the systematic 
errors arising from approximating the true signal with an imperfect theoretical 
template~\cite{PhysRevD.89.104023}.

%%%%%%%%%%%%%%%%%%%%%%%%
\subsubsection{Validation of the method: BH ringdown}
%%%%%%%%%%%%%%%%%%%%%%%%
As a check of our computation, we have reproduced the results of the BH 
ringdown 
analysis performed in Ref.~\cite{Berti:2005ys}. This can be achieved by 
neglecting the echo part of our template, i.e. by setting ${\cal R}=0$ or 
simply considering only the first term in Eq.~\eqref{signalomega}.

We have first reproduced the analytical results presented in Appendix~B 
of Ref.~\cite{Berti:2005ys} for what concerns the statistical errors on the 
parameters of the ringdown waveform. We have then derived the same results 
through a numerical integration of Eq.~\eqref{fisher}.
Note that Ref.~\cite{Berti:2005ys} adopted a $\delta$-approximation (i.e., 
white 
noise,
$S_n(f) = {\rm const}$), which is equivalent to consider the signal as 
monochromatic.  In this limit, the quantity $\rho \sigma_i$ is by 
construction 
independent of the detector sensitivity. As a 
further check, we have relaxed the $\delta$-approximation and repeated the 
analysis using the recently proposed LISA's noise spectral 
density~\cite{LISA}, obtaining similar results.

%%%%%%%%%%%%%%%%%%%%%%%%
\subsubsection{Analysis for GW echoes}
%%%%%%%%%%%%%%%%%%%%%%%%

After having validated our scheme, we computed numerically the Fisher 
matrix~\eqref{fisher} with the template~\eqref{FINALTEMPLATE} using the 
sensitivity curves presented in Fig.~\ref{fig:SNR} for current and future 
detectors. As previously discussed, for linearly polarized waves the template 
contains $5$ ringdown parameters (mass, spin, phase, amplitude, and starting 
time), and two ECO parameters (the frequency-dependent reflection 
coefficient ${\cal R}(\omega)$ and the width of the cavity $d$) which 
fully characterize the echoes. The parameter $d$ is directly related to 
physical quantities, such as the compactness of the ECO or the redshift at 
the surface.
%%

%%%%%%%%%%%%%%%%%%%%%%%%%%%%%%%%%%%%%%%%%%%%%%%%
\subsection{Results}
%%%%%%%%%%%%%%%%%%%%%%%%%%%%%%%%%%%%%%%%%%%%%%%%
%%%%
%
\begin{figure*}[th]
\centering
\includegraphics[width=0.47\textwidth]{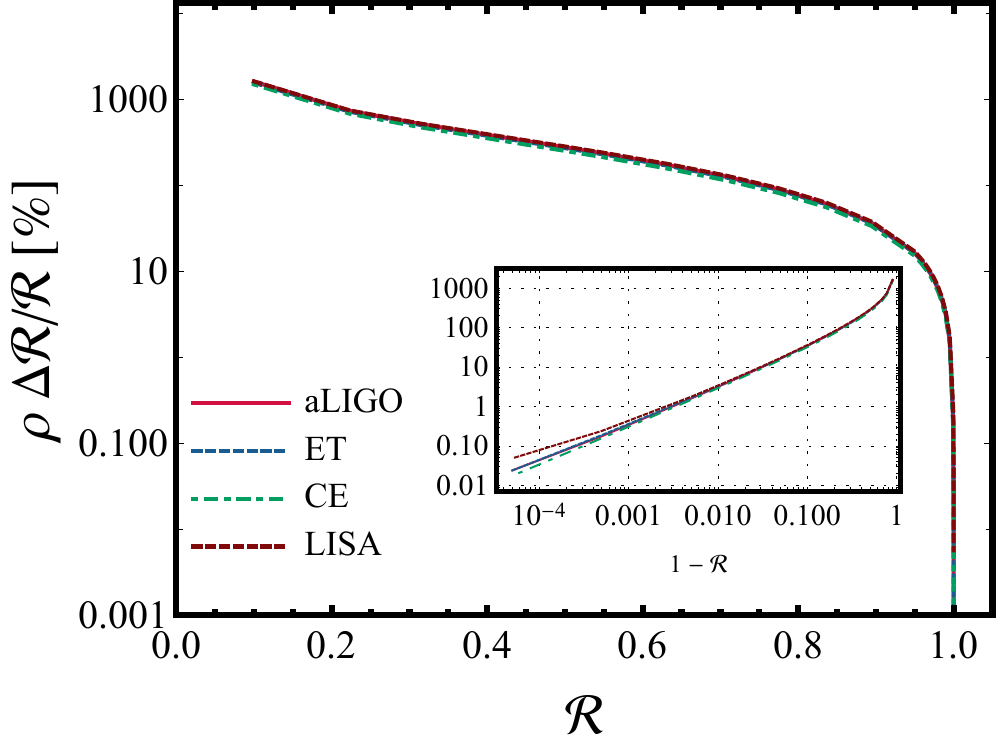}
\includegraphics[width=0.47\textwidth]{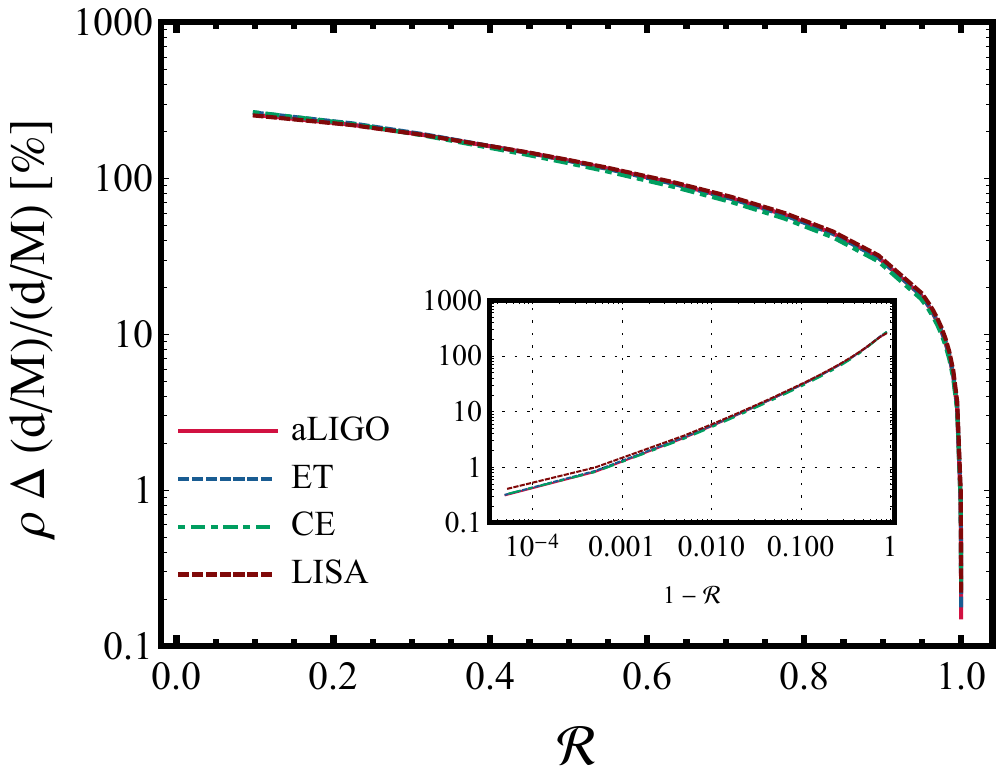}
\caption{Left panel: relative (percentage) error on the reflection coefficient, 
$\Delta {\cal R} / {\cal R}$ multiplied by the SNR, as a function of ${\cal 
R}$. The inset shows the 
same quantity as a function of $1-{\cal R}$ in a logarithmic scale.
Right panel: same as 
in the left panel but for the width of the cavity, 
$\Delta (d/M)/(d/M)$. For simplicity, we assumed that ${\cal R}$ 
is real and positive. 
In both panels we considered $M=30M_\odot$ for ground-based detectors and 
$M=10^6 M_\odot$ for LISA. We assume $d=100 M$ but the errors are independent 
of 
$d$ in the large-$d$ limit, see Appendix~\ref{app:d}.} 
\label{fig:errors}
\end{figure*}
%
%%%%%%
Our main results for the statistical errors on the ECO parameters are 
shown in Fig.~\ref{fig:errors}. In the large SNR limit, the errors scale as 
$1/\rho$ so we present the quantity $\rho\Delta {\cal R}/{\cal R}$ (left panel) 
and 
$\rho\Delta (d/M)/(d/M)$ (right panel\footnote{We 
adopted 
dimensionless parameters to define the Fisher matrix, in particular $M/M_\odot$ 
and $d/M$. The 
statistical error on $d$ can be easily obtained from the full 
covariance matrix.}). Figure~\ref{fig:errors} shows a number of 
interesting features:
%%%
\begin{itemize}
 \item The relative errors are almost independent of the sensitivity curve 
of the detector and only depend on the SNR. This suggests that the parameter 
estimation of echoes will be only mildly sensitive to the details of future 
detectors. Obviously, future interferometers on Earth and 
in space will allow for very high SNR in the post-merger phase 
$(\rho\approx 100$ for ET/Cosmic Explorer, and possibly even larger for LISA), 
which 
will put more stringent constraints on the ECO parameters.
This mild dependence of the relative errors on the sensitivity curve is valid 
for signals located near the minimum of the sensitivity curve, as those adopted 
in Fig.~\ref{fig:errors}. Less optimal choices of the injected parameters
would give a more pronounced 
(although anyway small) dependence, which is due to the different behavior of 
the various 
sensitivity curves at low/high frequencies.

%%%
\item Although Fig.~\ref{fig:errors} was obtained by injecting the 
value $d=100M$, the statistical absolute errors on ${\cal R}$ and $d$ do
not depend on 
the injected value of $d$ in the $d\to \infty$ limit. We give an analytical 
explanation of this seemingly counter-intuitive property in 
Appendix~\ref{app:d}. The statistical errors for $d=100M$ are very similar to 
those for $d=50M$ and saturate for larger values of $d/M$. Overall, our 
analysis suggests that the detectability of the echoes is independent of $d$ in 
the large-$d$ limit (i.e., for ultracompact objects).
%%%%
\item A further important feature is 
the strong dependence of the relative errors on the value of the reflection 
coefficient ${\cal R}$. In particular, the relative errors for ${\cal R}=1$ are 
smaller than those for ${\cal R}\approx0.5$ roughly by $4$ orders of 
magnitude. The 
reason for this is related to what is shown in Figs.~\ref{fig:SNR} 
and~\ref{fig:EvsR}: the amplitude of individual 
echoes and the total energy of the signal depend strongly on the reflection 
coefficient. This feature suggests that it should be relatively straightforward 
to rule out or detect models with ${\cal R}\approx 1$, whereas it is 
increasingly more difficult to constrain models with smaller values of ${\cal 
R}$.
\end{itemize}
%%%

As an example, let us consider the extremal case ${\cal R}=1$. Although this 
case is ruled out by the ergoregion 
instability~\cite{Cardoso:2008kj,Maggio:2017ivp} and by the 
absence of GW stochastic background in LIGO O1~\cite{Barausse:2018vdb}, it is 
interesting to explore the level of constraints achievable in this case. For 
a reference event with $M=30 M_\odot$ and $d>50 M$ with aLIGO, 
we obtain
%%%
\begin{equation}\label{estimateDeltaR}
 \frac{\Delta {\cal R}}{{\cal R}}\approx 
5\times 10^{-8}\left(\frac{8}{\rho_{\rm ringdown}}\right) \qquad {\rm for}\;\,	
{\cal R}=1\,,
\end{equation}
%%%
where, as a reference, we normalized $\rho$ such that the value of the SNR 
\emph{in the ringdown phase only}\footnote{When ${\cal R}\approx1$, the SNR in 
the ringdown phase can differ significantly from the 
SNR in the whole post-merger phase, 
see~Fig.~\ref{fig:EvsR}. For ${\cal R}=1$, and for the parameters considered in 
Eq.~\eqref{estimateDeltaR}, the total SNR is $\rho\approx 18\rho_{\rm 
ringdown}$.} is that of 
GW150914~\cite{GW150914}. This suggests that this model could be detected or 
ruled out compared to the BH case (${\cal R}=0$) at more than $5\sigma$ 
confidence level with aLIGO/Virgo.

Figure~\ref{fig:errors} also shows a strong dependence of $\Delta{\cal 
R}/{\cal R}$ when ${\cal R}<1$. It is therefore
interesting to calculate the SNR necessary to discriminate a 
partially-absorbing ECO from a BH on the basis of a measurement of ${\cal R}$ 
at some confidence level. Clearly, if $\Delta {\cal R}/{\cal R}>100\%$, the 
measurement would be compatible with the BH case (${\cal R}=0$). On the other 
hand, if $\Delta {\cal R}/{\cal R}<(4.5,0.27,0.007,0.00006)\%$, one might be 
able (by performing a more sophisticated analysis than the one presented here)
to detect or rule out a given model at $(2,3,4,5)\sigma$ confidence level, 
respectively.

%%%%%%%%%%%%%%%%%%%%%%%%
\begin{figure}[th]
\centering
\includegraphics[width=0.52\textwidth]{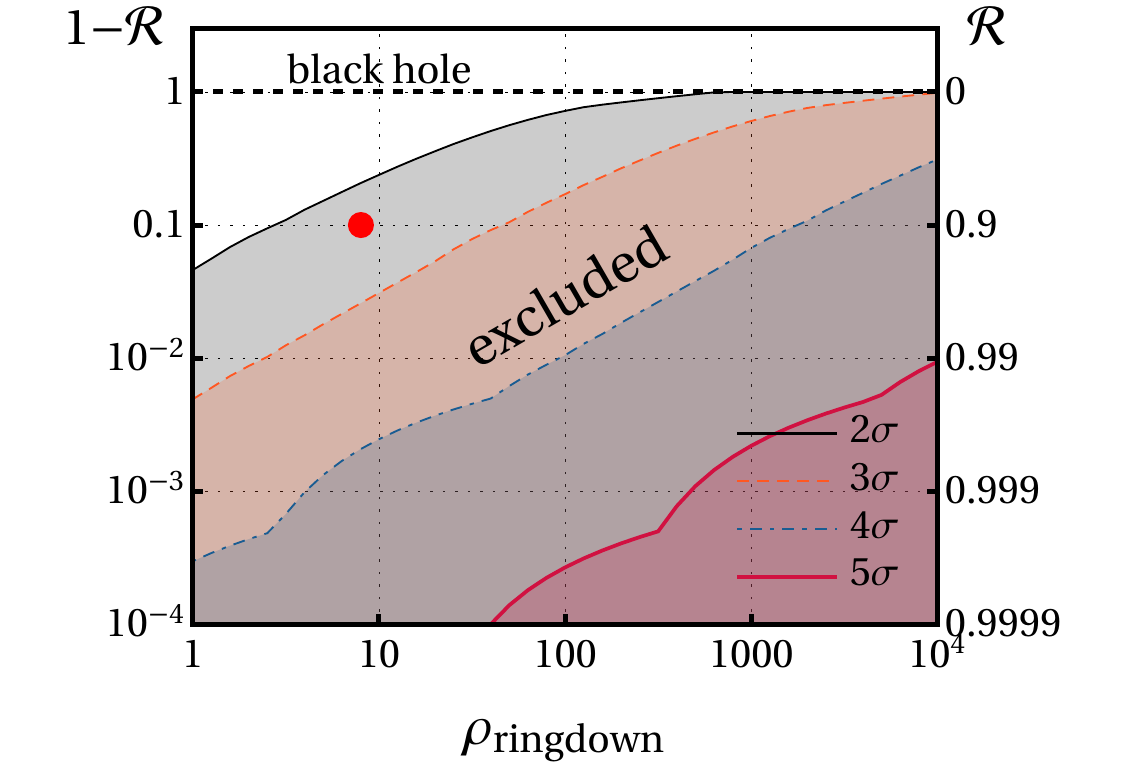}
\caption{Projected exclusion plot for the ECO reflectivity ${\cal R}$ as a 
function of the SNR in the ringdown phase and at different $\sigma$ 
confidence levels. 
The shaded areas represent regions that can be excluded at a 
given confidence level. 
This plot is based on Fig.~\ref{fig:errors} and assumes $d\gg M$ and $M= 
30M_\odot$ ($M=10^6M_\odot$) for ground- (space-) based detectors.
The red marker corresponds to $\rho_{\rm ringdown}=8$ and ${\cal 
R}=0.9$, which is the value claimed in Ref.~\cite{Abedi:2016hgu} at 
$(1.6\div2)\sigma$ level.} \label{fig:RvsSNR}
\end{figure}
%%%%%%%%%%%%%%%%%%%%%%%%

The result of this preliminary analysis is shown in Fig.~\ref{fig:RvsSNR}, 
where we present the exclusion plot for the parameter ${\cal R}$ as a function 
of the SNR in the ringdown phase. 
Shaded areas represent regions which can be 
excluded at some given confidence level. Obviously, larger SNRs would allow to 
probe values of ${\cal R}$ close to the BH limit, ${\cal R}\approx 0$.
The extent of the constraints strongly depends on the confidence 
level.
For example, ${\rm SNR\approx100}$ in the ringdown would allow to distinguish 
ECOs with ${\cal  R}\gtrsim 0.3$ from BHs at $2\sigma$ confidence level, but a 
$3\sigma$ detection would require ${\cal R}\gtrsim0.85$. The reason for this is 
again related to the strong dependence of the echo signal on ${\cal R}$ (see 
Figs.~\ref{fig:SNR} and~\ref{fig:EvsR}).
%%%

Finally, the red marker in Fig.~\ref{fig:RvsSNR} corresponds to a 
detection at $\rho=8$ in the ringdown phase (such as 
GW150914~\cite{GW150914}) and an ECO with reflectivity ${\cal 
R}=0.9$. These are roughly the values of the tentative detection claimed in 
Ref.~\cite{Abedi:2016hgu} at $(1.6\div2)\sigma$ level (but see also 
Refs.~\cite{Conklin:2017lwb,Ashton:2016xff,Abedi:2017isz,Westerweck:2017hus,
Abedi:2018pst}). Although our analysis is preliminary, it is interesting 
to note that our results are not in tension with the claim of 
Ref.~\cite{Abedi:2016hgu}, since Fig.~\ref{fig:RvsSNR} suggests that an ECO 
with 
${\cal R}=0.9$ could be detected at $\lesssim 2.5\sigma$ confidence level 
through an event with $\rho_{\rm ringdown}\approx8$.
On the other hand, the Fisher analysis only gives an estimate of the 
statistical errors based on a theoretical template, without using 
real data. As such, it would also be in 
agreement with a negative search or with smaller significance, like that 
reported in 
Ref.~\cite{Westerweck:2017hus}.
We consider the results 
shown in
Fig.~\ref{fig:RvsSNR} as merely indicative that interesting constraints on (or 
detection of) quantum-dressed ECOs are within reach of current and, 
especially,
future detectors.
Indeed, Fig.~\ref{fig:RvsSNR} shows that to confirm the putative detection of 
Ref.~\cite{Abedi:2016hgu} at $3\sigma$ ($4\sigma$) level would require a 
single-event detection with $\rho_{\rm ringdown}\approx 50$ ($\rho_{\rm 
ringdown}\approx1800$).

%%%%%%%%%%%%%%%%%%%%%%%%%
\section{Discussion and Outlook}~\label{sec:Discussion}
%%%%%%%%%%%%%%%%%%%%%%%%%

We have presented an analytical template that describes the ringdown and 
subsequent echo signal of a ultracompact horizonless object motivated by 
putative near-horizon quantum structures. This template 
depends on the physical parameters of the echoing remnant, such as the 
reflection coefficient ${\cal R}$ and the redshift at the surface of the object.
This study is the first step in the development of an accurate template to be 
used in direct searches for GW echoes using matched filters and in parameter 
estimation.

We have characterized some of the features of the template, which are anchored 
to the physical properties of the ECO model. The time-domain 
waveform contains all features previously reported for the echo signal, namely 
amplitude and frequency modulation, and phase inversion of each echo relative 
to 
the previous one due to the reflective boundary conditions.
The amplitude of subsequent echoes (both in the frequency and in the time 
domain) 
depends strongly on the reflectivity ${\cal R}$. When ${\cal R}\approx 
1$ 
the echo signal has amplitude and energy significantly larger than those 
of the ordinary BH ringdown. This suggests that GW echoes in 
certain models might be detectable even when the ringdown is not.

Using a Fisher analysis, we have then estimated the statistical errors on the 
template parameters for a post-merger GW detection with current and future 
interferometers. Interestingly, for signals in the optimal frequency window, 
the statistical errors at a given SNR depend only mildly on the detector's 
sensitivity curve.

Our analysis suggests that ECO models 
with ${\cal R}\approx 1$ can be detected or ruled out even with aLIGO/Virgo 
(for events with $\rho\gtrsim 8$ in the ringdown phase) at $5\sigma$ 
confidence level. The same event might allow us to probe values of 
the reflectivity as small as ${\cal R}\approx 0.8$ roughly at $2\sigma$ 
confidence level.
Overall, the detectability of the echoes is independent of the parameter $d$ in 
the large-$d$ limit (i.e., for ultracompact objects).

ECOs with ${\cal R}=1$ are ruled out by the ergoregion 
instability~\cite{Cardoso:2008kj,Maggio:2017ivp} and by the 
absence of GW stochastic background in LIGO O1~\cite{Barausse:2018vdb}. 
Excluding/detecting echoes for models with smaller values of the reflectivity 
(for which 
the ergoregion instability is absent~\cite{Maggio:2017ivp})
will require 
SNRs in the post-merger phase of ${\cal O}(100)$. This will be achievable only 
with third-generation detectors (ET and Cosmic Explorer) and with the 
space-mission LISA.

Although our analysis is preliminary, we believe that our results 
already indicate that interesting constraints on (or 
detection of) quantum-dressed ECOs are within reach with current (and 
especially 
future) interferometers.

Extensions of this work are manifold. A template valid for the spinning case 
is underway. This case is particularly interesting, not only because merger 
remnants are spinning, but also because of the rich 
phenomenology of spinning horizonless objects, which might undergo various 
types 
of instabilities~\cite{1978CMaPh..63..243F,Moschidis:2016zjy,Brito:2015oca,
Cardoso:2007az,Cardoso:2008kj,Cardoso:2014sna,Maggio:2017ivp}. In particular, 
due to superradiance~\cite{Brito:2015oca} and to the ergoregion 
instability~\cite{Cardoso:2014sna,Vicente:2018mxl}, the echo signal will grow 
in time over a timescale $\tau_{\rm ergoregion}$ which is generically much 
longer than $\tau_{\rm echo}\sim d$.
Another natural extension concerns the role of the boundary conditions at the 
surface for gravitational perturbations~\cite{Price:2017cjr} --~especially in 
the spinning case~-- and on model-independent ways to describe the interior of 
the object in case of partial absorption (cf., e.g., 
Ref.~\cite{Barausse:2018vdb}).

Furthermore, a more realistic model could be obtained by reprocessing the full 
form of $Z_{\rm BH}^+$ containing both the ringdown and the late-merger signal 
[i.e., using the template~\eqref{FINALTEMPLATEGEN}], or using a superposition 
of QNMs. This extension will be particularly 
important to compare our template (constructed within perturbation theory and 
therefore strictly speaking valid only for weak sources) with the post-merger 
signal of 
coalescences forming an ``echoing'' ultracompact horizonless object. 
Unfortunately, numerical simulations of these systems are 
currently unavailable, but we envisage that a comparison between analytical 
and numerical waveforms will eventually follow a path similar to what done in 
the past for the matching of ringdown templates with numerical-relativity 
waveforms (see, e.g., Ref.~\cite{Buonanno:2006ui}).

Another interesting prospect is to analyze the prompt-ringdown 
signal and the late-time one separately. Since the prompt response is 
universal~\cite{Cardoso:2016rao}, it could be used to 
infer the mass and the spin of the final object, thus making it easier to 
extract the echo parameters from the post-merger phase at late times.
Finally, it should be straightforward to extend our analysis to complex 
and (possibly) frequency-dependent values of ${\cal R}$ and to include two 
polarizations.
These analyses and other applications are left for future work.

%%%%%%%%%%%%%%%%%%%%%%%%%%%%%%%%%%%%%%%%%%%%%%%%%%%%%%%%%%%%%%%%%%%%%%
\begin{acknowledgments}
%%%%%%%%%%%%%%%%%%%%%%%%%%%%%%%%%%%%%%%%%%%%%%%%%%%%%%%%%%%%%%%%%%%%%%%
The authors acknowledge interesting discussion with the members of the 
GWIC-3G-XG/NRAR working group, the financial support provided under the 
European 
Union's H2020 ERC, Starting Grant agreement no.~DarkGRA--757480, and networking 
support by the COST Action CA16104. PP acknowledges the kind hospitality of the
Universitat de les Illes Balears, where this work has been finalized.
\end{acknowledgments}
%%%%%%%%%%%%%%%%%%%%%%%%%%%%%%%%%%%%%%%%%%%%%%%%%%%%%%%%%%%%%%%%%%%%%%%%%%%%%%
% 

\appendix

%%%%%%%%%%%%%%%%%%%%%%%%%%%%%%%%%%%%%%%%%%%%%%%%%%%%%%%%%%%%%%%%%%%%%%%%%%%%%%
\section{Transfer function in the geometrical optics approximation} 
\label{app:optics}
%%%%%%%%%%%%%%%%%%%%%%%%%%%%%%%%%%%%%%%%%%%%%%%%%%%%%%%%%%%%%%%%%%%%%%%%%%%%%%
In this appendix we derive Eqs.~\eqref{signalomega} and 
\eqref{eq:bhtemplateHOR} 
using a geometrical optics analogy. 
For simplicity, we fix the position of the peak of the potential at $x = 0$ 
rather than at $x=x_m$.
We replace the reflective surface and the light ring with two partially 
reflective mirrors, respectively left (L) at $x=- d < 0$ and right (R) at $x= 
0$, and each Fourier 
component of the perturbation with a light ray represented by a complex number. 
We are left with a Fabry-Perot interferometer that can be studied as in 
optics textbooks.
In the main text we consider a source $\tilde S(\omega,x)=C(\omega) \delta(x - 
x_s)$, which is symmetric with respect to the direction of propagation.
This will produce two rays originating at $x_s = -d  + {\ell}$, one going 
to 
the left and the other to the right with the same amplitude $C(\omega)$.
The left-going ray travels toward L acquiring a phase $e^{i\omega {\ell}}$, 
is reflected at L gaining a factor $\cal{R}$, and then acquires an 
extra phase
$e^{i\omega {\ell}}$ before joining the first ray in their common future 
evolution. The two rays then travel toward R with a total amplitude 
$C(\omega)(1 + {\cal R} e^{2i \omega {\ell}})$. Travelling to R they 
gather a
phase $e^{i\omega (d - {\ell})}$. Upon impinging on R the wave is partly 
reflected and partly transmitted. The transmitted part gains a factor ${\cal 
T}_{\rm BH}(\omega)$ and then travels to $+ \infty$ as
%%%
\begin{eqnarray}
&\sim& C(\omega) e^{ i\omega (d - {\ell})} 
{\cal 
T}_{\rm BH}(\omega) (1 + {\cal R} e^{2i \omega {\ell}}) e^{i \omega x} 
\nonumber \\
 &=&\tilde Z_{\rm BH}^{+}(\omega) (1 + {\cal R} e^{2i \omega {\ell}}) e^{i 
\omega x}\,, \label{totalrightwave}
\end{eqnarray}
where
\begin{eqnarray}\label{zinfApp}
\tilde Z_{\rm BH}^{+}(\omega) = C(\omega) e^{i \omega (d - {\ell})} {\cal 
T}_{\rm BH}(\omega)
\end{eqnarray}
is the BH response, corresponding to the amplitude of the right-going wave in 
Eq.~\eqref{totalrightwave} when ${\cal R}=0$.
On the other hand, the reflected wave gains a 
factor ${\cal R}_{\rm BH}(\omega)$ and travels from R to L, acquiring an extra 
phase $e^{i \omega d}$, and keeps repeating the same steps already described.
%

%%%

To compute $\tilde Z_{\rm BH}^{-}$ we consider the amplitude of the total 
left-going signal near $x\sim -d$ and set 
${\cal R} = 0$:
\begin{equation}\label{zhApp}
\tilde Z_{\rm BH}^{-} = \left[C(\omega) e^{i\omega {\ell}}+ C(\omega) {\cal 
R}_{\rm BH}(\omega) e^{ 2 i \omega d} e^{- i\omega {\ell}} \right] e^{-i 
\omega 
d}\,,
\end{equation}
where we added the phase factor $e^{-i \omega d}$ to take into account the 
convention about the origin of the asymptotic behaviors.
Considering Eqs.~\eqref{zinfApp} and \eqref{zhApp} we recover 
Eq.~\eqref{eq:bhtemplateHOR} for $x_m = 0$ (i.e. $x_0=-d$).

From Eqs.~\eqref{totalrightwave} and \eqref{zhApp}, it is now straightforward to calculate the signal at infinity as a sum of the 
transmitted waves at R:
%%%
\begin{eqnarray}
 \tilde Z^{+}(\omega) &=& (1 + {\cal R} e^{2i \omega {\ell}}) Z^{+}_{\rm 
BH}(\omega) \left[ 1+ \sum_{n = 1}^{\infty} ({\cal R} {\cal R}_{\rm BH} e^{2i
\omega d})^n \right] \nonumber \\
&=&(1 + {\cal R} e^{2i \omega {\ell}}) Z^{+}_{\rm BH}(\omega) \frac{1}{1 - 
{\cal R} {\cal R}_{\rm BH} e^{2i \omega d}}\,,\label{cong}
\end{eqnarray}
%%%%
which gives Eqs.~\eqref{signalomega} and \eqref{transfer} when $\tilde Z_{\rm 
BH}^-$ is given by Eq.~\eqref{eq:bhtemplateHOR}.
%fwith
The above argument can also be extended to generic sources~\cite{thesis}.

%
%%%%%%%%%%%%%%%%%%%%%%%%%%%%%%%%%%%%%%%%%%%%%%%%%%%%%%%%%%%%%%%%%%%%%%%%%%%%%%
\section{The functions ${\cal T}_{\rm BH}$, ${\cal R}_{\rm BH}$ and ${\cal K}$ 
in the analytical approximation} \label{app:comparison}
%%%%%%%%%%%%%%%%%%%%%%%%%%%%%%%%%%%%%%%%%%%%%%%%%%%%%%%%%%%%%%%%%%%%%%%%%%%%%%

Here we compare our approximate analytical results with the exact numerical 
ones as computed by Mark 
et al.~\cite{Mark:2017dnq}. Note that our definition of the tortoise coordinate 
$x$ differs by a constant term, $-2M\log2$, relative to the one adopted in 
Ref.~\cite{Mark:2017dnq}. For the purpose of comparison, only in this section 
we 
have rescaled $x_0\to x_0-2M\log2$ to agree with the definition of 
Ref.~\cite{Mark:2017dnq}. 
%%%

\begin{figure*}[th]
\centering
\includegraphics[width=0.46\textwidth]{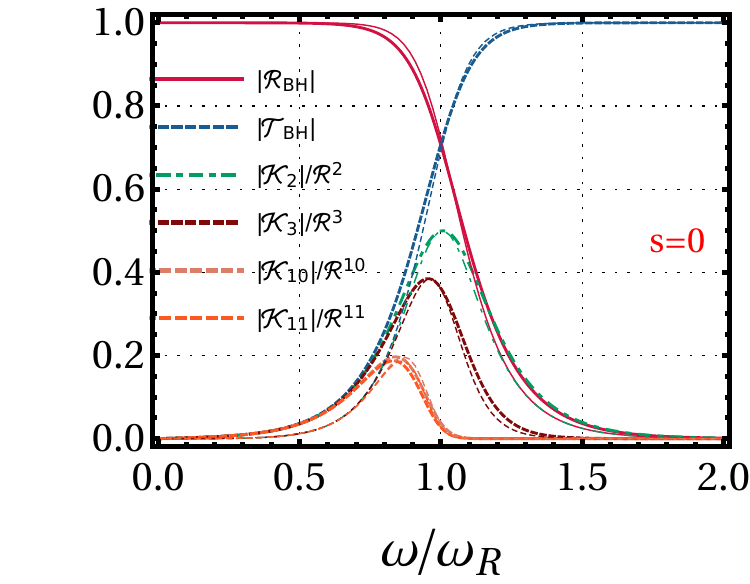}
\includegraphics[width=0.49\textwidth]{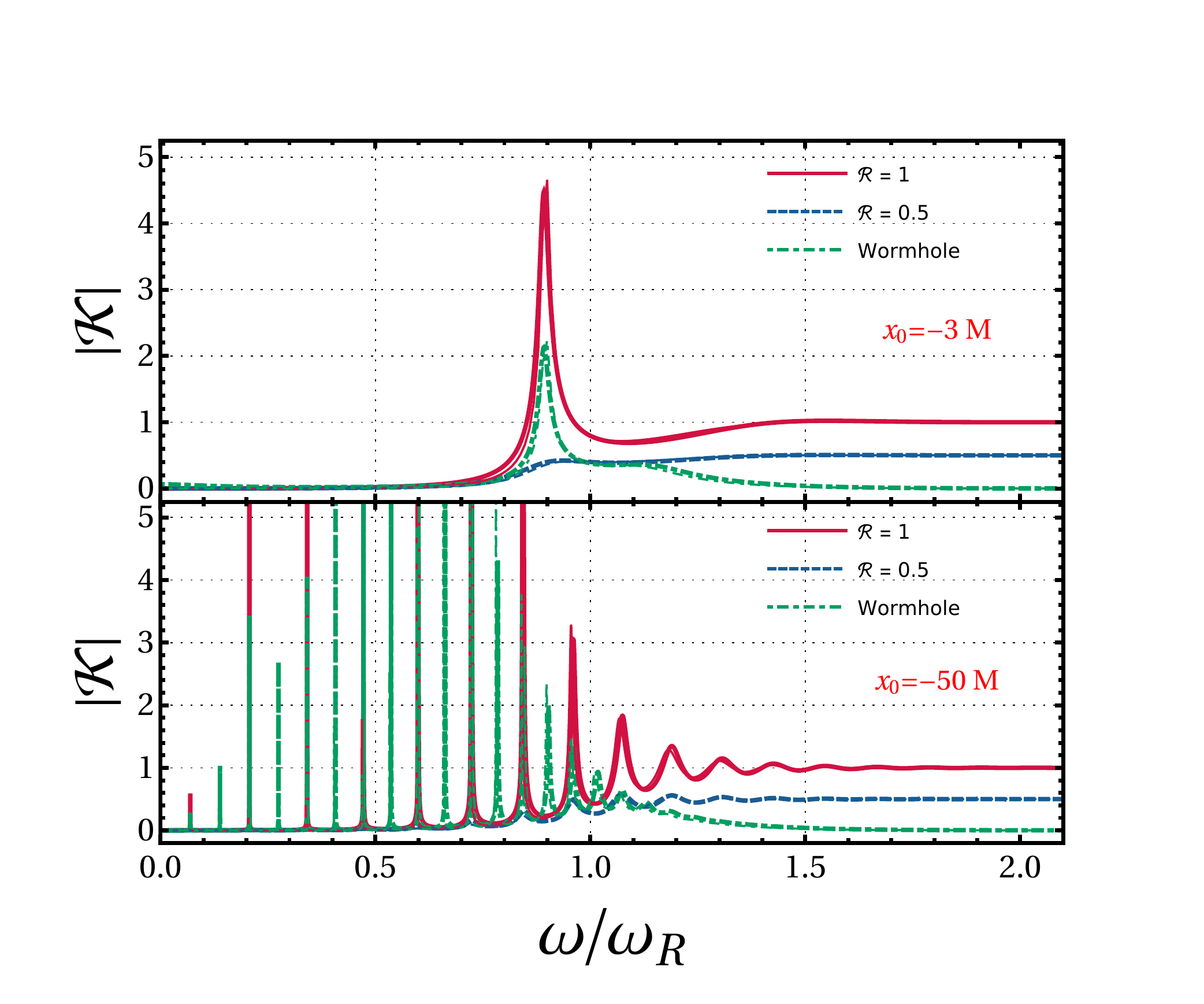}
\caption{Left panel: the functions ${\cal T}_{\rm BH}(\omega)$ and ${\cal 
R}_{\rm BH}(\omega)$ as computed in our analytical approximation, together with 
the functions ${\cal K}_n(\omega)$ defined in Eq.~\eqref{Knomega}. Thin curves 
correspond to the exact numerical results with the same color and linestyle. 
Right panel: same as the left panel but for the transfer function ${\cal 
K}(\omega)$ in various cases. Note that the agreement for the full transfer 
function ${\cal K}(\omega)$ is better than that for the individual ${\cal 
K}_n$.} 
\label{fig:TBK}
\end{figure*}

In the left panel of Fig.~\ref{fig:TBK} we compare the approximate analytical 
functions 
${\cal T}_{\rm BH}(\omega)$, ${\cal R}_{\rm BH}(\omega)$ with their exact
numerical behavior. This plot reproduces Fig.~3 in 
Ref.~\cite{Mark:2017dnq}. We also compare the 
functions
\begin{equation}
{\cal K}_n(\omega) = ({\cal T}_{\rm BH} {\cal R}) ({\cal R}_{\rm BH} {\cal 
R})^{n - 1} e^{- 2 i \omega x_0}\,, \label{Knomega}
\end{equation}
%%%
for different values of $n$ with their numerical counterparts. Note that we 
show 
the absolute value of ${\cal K}_n$ and normalize it by ${\cal R}^n$; with this 
choice, the result is independent of $x_0$ and ${\cal R}$.

In the right panel of 
Fig.~\ref{fig:TBK} we present the analytical transfer function ${\cal 
K}(\omega)$ [cf. Eq.~\eqref{Kappa}] for different values of $x_0$ and ${\cal 
R}$, and compare it with its exact numerical expression. This plot reproduces
Fig.~13 in Ref.~\cite{Mark:2017dnq}. We note that, while the agreement of the 
analytical approximation for the full 
transfer function ${\cal K}(\omega)$ is very good (the differences between 
numerics and analytics are barely distinguishable in the right panel of 
Fig.~\ref{fig:TBK}), the single terms ${\cal 
K}_n(\omega)$ are less accurate.

Note also the appearance of more 
narrow resonances as $|x_0|\gg M$, some of them not being resolved in Fig.~13 
of 
Ref.~\cite{Mark:2017dnq}. These resonances correspond to the long-lived 
QNMs of the ECO: their separation is $\sim \pi/d$ and their narrow width is 
associated with the small imaginary part of the quasi-bound 
modes~\cite{Cardoso:2014sna,Brito:2015oca}.

For example, Eq.~\eqref{eq:boundary} implies Dirichlet or Neumann boundary 
conditions at $x=x_0$ for ${\cal R}=-1$ or ${\cal R}=1$, respectively. In this 
case, the normal modes of a cavity of width $d$ at low frequency (${\cal T}_{\rm 
BH}\approx0$) read
%%%
\begin{equation}
 \omega_D \sim \frac{p\pi}{d}\,,\qquad \omega_N \sim \frac{(2p+1)\pi}{2d}\,, 
\label{modes}
\end{equation}
%%%
for Dirichlet and Neumann boundary conditions, respectively, and where $p$ 
is an integer.

In the right panel of Fig.~\ref{fig:TBK} we also show the case of a 
wormhole, for which ${\cal R}={\cal R}_{\rm BH} e^{-2i\omega 
x_0}$~\cite{Mark:2017dnq}. Although marginally visible in Fig.~\ref{fig:TBK}, 
in this case ${\cal K}(\omega)$ does not vanish as $\omega\to0$. This behavior 
holds generically whenever ${\cal R}(\omega \approx 0) = -1 + {\cal 
O}(\omega)$. Indeed, in this case we get
%%%
\begin{equation}\label{eq:kappa0}
{\cal K}(0)=\frac{\pi  \sech (\frac{\pi \omega_R}{\alpha} )}{H_{-i \frac{\omega_R}{\alpha}-\frac{1}{2}}+H_{i \frac{\omega_R}{\alpha}-\frac{1}{2}}-\alpha(2 d + i {\cal R}'(0) )}\,,
\end{equation}
with $H_n = \digamma(n + 1) + \gamma_E $, $\digamma$ is the digamma function and $\gamma_E$ is the Euler-Mascheroni constant.
%%%
As expected ${\cal K}(0) \to 0$ as $d \to \infty$ (assuming no cancellations occur due to the ${\cal R}'(0)$ term) but is otherwise finite. 
By differentiating ${\cal R}(\omega) = {\cal R}^*(-\omega)$ we see that ${\cal 
R}'(0)$ is purely imaginary, and therefore ${\cal K}(0)$ is purely real, in 
accordance with Eq.~\eqref{eq:kreal}.
In the wormhole case, ${\cal R}'_{\rm WH}(0) =  {\cal R}'_{\rm BH}(0)- 2 i d $, and we obtain
\begin{equation}
{\cal K}(0)=\frac{  \pi\sech (\frac{\pi \omega_R}{\alpha})/2}{H_{-i \frac{\omega_R}{\alpha}-\frac{1}{2}}+H_{i \frac{\omega_R}{\alpha}-\frac{1}{2}}-2 \alpha  d}\,.
\end{equation}
%

%%%%%%%%%%%%%%%%%%%%%%%%%%%%%%%%%%%%%%%%%%%%%%%%%%%%%%%%%%%%%%%%%%%%%%%%%%%%%%
\section{Frequency-domain analytical template for a source localized in a 
generic position} 
\label{app:genericsource}
%%%%%%%%%%%%%%%%%%%%%%%%%%%%%%%%%%%%%%%%%%%%%%%%%%%%%%%%%%%%%%%%%%%%%%%%%%%%%%
Equation~\eqref{FINALTEMPLATEGEN} in the main text was derived assuming a source 
localized near the surface at $x_s\approx x_0$. It is possible to extend this 
result to a source localized at any $x_s$~\cite{thesis}. From the analytical 
solution of the homogeneous problem [Eq.~\eqref{tildePsi0}], it is possible to 
derive the two independent solutions $\tilde\Psi_\pm$ that appear in the 
definition~\eqref{ZBH}. If $\tilde S=C(\omega)\delta (x-x_s)$, then
%%%%
\begin{equation}
 Z_{\rm BH}^-(\omega) = Z_{\rm 
BH}^+(\omega)\frac{\tilde\Psi_+(x=x_s,\omega)}{\tilde\Psi_-(x=x_s,\omega)}\,,
\end{equation}
%%%%
which is again valid for any function $C(\omega)$ characterizing the source.
It is now straightforward to obtain the generalization of 
Eq.~\eqref{FINALTEMPLATEGEN}:
%%%
 \begin{equation}
\tilde Z^+(\omega) = Z_{\rm BH}^+(\omega) \left(1-{\cal G} {\cal R}e^{2i\omega 
d} \right) \,, 
\label{FINALTEMPLATEGENSOURCE}
 \end{equation}
where 
\begin{widetext} 
\begin{equation}
 {\cal G} = \frac{\pi \sinh\left[\frac{\pi\omega}{\alpha}\right]\Upsilon 
{P_{\frac{i {\omega_R}}{\alpha 
}-\frac{1}{2}}^{\frac{i \omega}{\alpha }}(y)}}{\left(\pi +e^{2 i d \omega } 
{\cal R}
   {\cosh}\left[\frac{\pi  {\omega_R}}{\alpha }\right]
   \Upsilon\right) \left(\pi {\sinh}\left[\frac{\pi  
(\omega+{\omega_R})}{\alpha }\right]{P_{\frac{i {\omega_R}}{\alpha 
}-\frac{1}{2}}^{\frac{i \omega}{\alpha }}(y)}-2 i
   {\cosh}\left[\frac{\pi  (\omega +{\omega_R})}{\alpha }\right]
   {Q_{\frac{i {\omega_R}}{\alpha 
}-\frac{1}{2}}^{\frac{i \omega}{\alpha }}(y)}\right)}\,,
\end{equation}
\end{widetext}
%%%
$\Upsilon$ is defined in Eq.~\eqref{gamma}, $y=-{\tanh}(\alpha  \Delta)$, and 
$\Delta=x_m-x_s$. Note that $\Delta=d$ when $x_s=x_0$ and that $\Delta>0$ for a 
source localized in the cavity, whereas $\Delta<0$ for a source localized 
outside the photon-sphere.
%%%
Equation~\eqref{FINALTEMPLATEGENSOURCE} reduces to 
Eq.~\eqref{FINALTEMPLATEGEN} when $\Delta =d\gg M$, as expected.

%%%%%%%%%%%%%%%%%%%%%%%%%%%%%%%%%%%%%%%%%%%%%%%%%%%%%%%%%%%%%%%%%%%%%%%%%%%%%%
\section{A toy model for the dependence on $d$ of the Fisher matrix} 
\label{app:d}
%%%%%%%%%%%%%%%%%%%%%%%%%%%%%%%%%%%%%%%%%%%%%%%%%%%%%%%%%%%%%%%%%%%%%%%%%%%%%%
In this appendix we give an analytical argument showing that the absolute 
errors on our template's parameters computed through the Fisher matrix are 
independent of $d$ in the large-$d$ limit.

We start by modeling the echo signal in the time domain as
\begin{equation}\label{eq:td}
\Psi(t) = \sum_{n = 0}^{\infty} g(t - 2 n d) e^{-n\gamma}\,,
\end{equation}
where $d$ is the width of the cavity, $\gamma$ is a dumping factor, and $g$ is 
a generic function.
Taking the Fourier transform of Eq.~\eqref{eq:td}, we obtain
\begin{equation}
\tilde \Psi(\omega) = \sum_{n = 0}^{\infty} e^{-n\gamma} \int_{-\infty}^{+ 
\infty}  \frac{dt}{\sqrt{2 \pi}}  g(t - 2 n d) e^{i \omega t}\,.
\end{equation}
%%%
By making the change of variable $t' = t -2 n d$, we get
\begin{eqnarray}
\tilde \Psi(\omega) &=& \sum_{n = 0}^{\infty} e^{-n\gamma} \int_{-\infty}^{+ 
\infty} \frac{dt'}{\sqrt{2 \pi}} g(t') e^{i \omega t'}e^{2 i \omega n d}  \nn\\
&=&  \sum_{n = 0}^{\infty} e^{-n \gamma} e^{2 i \omega n d}  \int_{-\infty}^{+ 
\infty} \frac{dt}{\sqrt{2 \pi}} g(t) e^{i \omega t}  \nn\\
&=& \tilde g(\omega) A(\omega; \gamma, d)\,,
\end{eqnarray}
where
\begin{equation}
A(\omega; \gamma, d) = \frac{1}{1 -e^{- (\gamma - 2 i \omega d)}}.
 \end{equation}
We notice that $A(\omega; \gamma, d)$ has the following properties
\begin{eqnarray}
\lim_{\frac{d}{d_\text{c}} \to \infty} \int_{-\infty}^{+\infty} 
f(\omega)A(\omega; \gamma, d) d\omega = c_1(\gamma) \int_{-\infty}^{+\infty} f(\omega) 
d\omega\,, \nonumber \\
\lim_{\frac{d}{d_\text{c}} \to \infty}  \frac{\partial}{\partial 
\gamma}\int_{-\infty}^{+\infty} f(\omega)A(\omega; \gamma, d) d\omega = c_2(\gamma) 
\int_{-\infty}^{+\infty} d\omega f(\omega)\,, \nonumber \\
\lim_{\frac{d}{d_\text{c}} \to \infty}  \frac{\partial}{\partial 
d}\int_{-\infty}^{+\infty} d\omega f(\omega)A(\omega; \gamma, d)  = c_3(\gamma) 
\int_{-\infty}^{+\infty} d\omega f(\omega) \omega\,,  \nonumber 
\end{eqnarray}
where $\sim 1/d_\text{c}$ is the characteristic scale of 
variations of the function $f(\omega)$. 
From the above relations, it follows immediately that the absolute statistical 
errors derived from the Fisher matrix are independent of $d$ when $d\gg d_c$.

In our specific case, the typical frequency scale of the function $\tilde 
g(\omega)$ is the BH ringdown frequency, $\omega_R\approx 0.37/M$. Thus the large-$d$ limit is achieved when
\begin{equation}
\frac{0.37}{M} \gg \frac{\pi}{d}\,,
\end{equation}
which requires $d\gg 10 M$. For larger values of $d$, we expect the 
relative errors such as $\Delta {\cal R}/{\cal R}$ to be independent of 
$d$, whereas $\Delta d/d\sim 1/d$. The numerical results presented in the main 
text support this expectation when $d\gg50M$.

% 
%%%%%%%%%%%%%%%%%%%%%%%%%%%%%%%%%%%%%%
% \bibliographystyle{apsrev4}
\bibliographystyle{utphys}
\bibliography{Ref}
\end{document}